\documentclass[aps,prx,11pt,amsmath,onecolumn,notitlepage,tightenlines,superscriptaddress,floatfix,eqsecnum]{revtex4}

\newcount\figstyle
\usepackage{amsmath}
\usepackage{amsfonts}
\usepackage{graphicx}
\usepackage{color}
\usepackage{epstopdf}
\usepackage{epsfig}
\usepackage{soul} 
\usepackage[breaklinks=true]{hyperref}
\usepackage{url}
\usepackage{yfonts}[1998/10/03]

\usepackage{tikz}

\newcommand{\sE}{\mathcal{E}}
\newcommand{\sB}{\mathcal{B}}
\newcommand{\sA}{\mathcal{A}}

\newcommand{\tr}{\mbox{tr}}
\newcommand{\dtwo}[1]{d^{\hspace{0.07em}2}\hspace{-0.08em}#1}

\newcommand{\mytilde}{\raise.17ex\hbox{$\scriptstyle\mathtt{\sim}$}}

\begin{document}

\title{Quantum limits on phase-preserving linear amplifiers}

\author{Carlton M. Caves}
\email{caves@info.phys.unm.edu}
\affiliation{Center for Quantum Information and Control, University of New Mexico, Albuquerque, NM 87131-0001, USA}
\affiliation{Centre for Engineered Quantum Systems, School of Mathematics and Physics,
University of Queensland, Brisbane, Queensland 4072, Australia}

\author{Joshua Combes}
\affiliation{Center for Quantum Information and Control, University of New Mexico, Albuquerque, NM 87131-0001, USA}

\author{Zhang Jiang}
\affiliation{Center for Quantum Information and Control, University of New Mexico, Albuquerque, NM 87131-0001, USA}

\author{Shashank Pandey}
\affiliation{Center for Quantum Information and Control, University of New Mexico, Albuquerque, NM 87131-0001, USA}

\begin{abstract}
The purpose of a phase-preserving linear amplifier is to make a small signal larger, regardless of its phase, so that it can be perceived by instruments incapable of resolving the original signal, while sacrificing as little as possible in signal-to-noise.  Quantum mechanics limits how well this can be done: a high-gain linear amplifier must degrade the signal-to-noise; the noise added by the amplifier, when referred to the input, must be at least half a quantum at the operating frequency.  This well-known quantum limit only constrains the second moments of the added noise.  Here we derive the quantum constraints on the entire distribution of added noise: we show that any phase-preserving linear amplifier is equivalent to a parametric amplifier with a physical state for the ancillary mode; the noise added to the amplified field mode is distributed according to the Wigner function of the ancilla~state.
\end{abstract}

\date{\today}

\maketitle

\section{Introduction}
\label{sec:intro}

The study of quantum limits on linear amplifiers became important in the 1960s with the invention and use of masers as microwave amplifiers.  Initial investigations~\cite{Louisell1961a,Heffner1962a,Haus1962a,Gordon1963b,Gordon1963a} led to the realization that quantum mechanics requires all phase-preserving linear amplifiers to add noise, thereby degrading the signal-to-noise ratio of the input signal.  For a high-gain linear amplifier, the minimum amount of added noise, when referred to the input of the amplifier, is equivalent to half a quantum at the operating frequency~\cite{Haus1962a,Caves1982a}.  For a quantum-limited input signal, this means a doubling of the input signal's zero-point noise and a halving of the input signal-to-noise ratio.

This fundamental quantum limit is expressed formally as a bound on the second moment of the noise added by a phase-preserving linear amplifier.   A comprehensive review article reprises the development and elaboration of this fundamental quantum limitation on the operation of linear amplifiers~\cite{Clerk2010a}.  In recent years, microwave-frequency amplifiers, based on the Josephson effect, have very closely approached the fundamental quantum limit on second-moment added noise~\cite{Bergeal2010b,Bergeal2010a,Kinion2011a}.  In the meantime, workers in quantum optics have formulated techniques for determining photon correlation functions without using photon counting, instead using the linear detection that at optical frequencies comes from homodyne detection~\cite{Grosse2007a}.  Researchers working with linear amplifiers at microwave frequencies have refined and elaborated these techniques into methods for determining the noise properties of signals input to a linear amplifier and of the added amplifier noise~\cite{daSilva2010a,Menzel2010a,Filippov2011a}.  These methods have been used to determine moments of amplifier noise well beyond second moments~\cite{Menzel2010a,Mariantoni2010a}, to measure photon correlation functions of input microwave signals~\cite{Bozyiget2010a}, to do quantum tomography on itinerant (wave-packet) microwave photons~\cite{Eichler2011b}, and to study squeezing of microwave fields~\cite{Eichler2011a,Mallet2011a}.

All these developments motivate an investigation of quantum limits on all moments of the added noise or, equivalently, on the entire distribution of added noise.  Second moments are sufficient to characterize the added noise if it is Gaussian; measuring higher moments allows one both to check the Gaussianity of the added noise and to characterize the performance of linear amplifiers more thoroughly.  Here we consider the case of phase-preserving amplification of a single bosonic mode, which we call the \textit{primary mode\/} and which has annihilation operator~$a$.  We characterize the input and output noise in terms of symmetrically ordered moments of $a$ and $a^\dagger$ or, equivalently, in terms of symmetrically ordered moments of the input and output quadrature components.  With this convention, the noise is described completely by the input and output Wigner functions of the primary mode.

We show here that regardless of how a phase-preserving linear amplifier is realized physically, it is equivalent to a parametric amplifier, i.e., an amplifier in which the primary mode undergoes a two-mode squeezing interaction with a single \textit{ancillary mode}, which has annihilation operator $b$.  The strength of the parametric interaction determines the amplifier's gain, and the noise added by the amplifier is distributed according to the Wigner function of the ancillary mode's initial state $\sigma$.  Characterizing completely the noise properties of a phase-preserving linear amplifier thus amounts to giving the initial state of this effective ancillary mode, even though the amplifier might be nothing like a parametric amplifier.  A quantum-limited (ideal) linear amplifier corresponds to the case where $\sigma$ is the vacuum state.

We begin by reviewing in Sec.~\ref{subsec:secmom} the simple input-output relation that leads to the second-moment constraint on added amplifier noise.  An ideal linear amplifier saturates the second-moment constraint and has Gaussian noise.  In Sec.~\ref{subsec:ideal} we give a stick-figure pictorial representation of the input and output noise in terms of contours of the popular quasidistributions for a field mode~\cite{Cahill1969a,Cahill1969b,Hillery1984a,GarrisonChiao2008}, the Glauber-Sudarshan $P$ function~\cite{Glauber1963a,Sudarshan1963a,Glauber1963b}, the Wigner function $W$~\cite{Wigner1932a}, and the Husimi $Q$ distribution~\cite{Husimi1940a}, and in Sec.~\ref{subsec:idealmodels} we consider four generic models for an ideal linear amplifier.  In Sec.~\ref{sec:char} we develop a general mathematical description of a linear amplifier that adds arbitrary noise.  The amplifier is described in terms of a linear map that takes the input state to the output state; this amplifier map must be completely positive~\cite{Kraus1983a,Nielsen2000a} to correspond to a physical linear amplifier.  Section~\ref{sec:genlim} formulates and proves our main result: the requirement of complete positivity implies that any linear amplifier is equivalent to a parametric amplifier with a physical initial state $\sigma$ for the ancillary mode.  Section~\ref{sec:examples} considers examples of nonideal amplifiers, including unphysical ones, and Sec.~\ref{sec:momlim} uses our main result to spell out the quantum limits on higher moments of the added noise.  Section~\ref{sec:con} sums up and briefly sketches future work.  The manipulations necessary to relate various kinds of moments are given in an Appendix.

\section{Quantum-limited phase-preserving linear amplifiers}
\label{sec:secmom}

\subsection{Quantum limit on second moment of added noise}
\label{subsec:secmom}

The setting for our investigation is a single bosonic mode, called the primary mode, which is to undergo phase-preserving linear amplification.  The primary mode has annihilation and creation operators,
\begin{align}
a&=\frac{1}{\sqrt2}(x_1+ix_2)\;,\\
a^\dagger&=\frac{1}{\sqrt2}(x_1-ix_2)\;;
\end{align}
in these expressions, the rapid oscillation at the modal frequency $\omega$ has been removed, and $x_1$ and $x_2$ are the Hermitian quadrature components of the mode.  The creation and annihilation operators obey the canonical commutation relation, $[a,a^\dagger]=1$ (equivalently, $[x_1,x_2]=i$).  This implies an uncertainty principle, $\langle\Delta x_1^2\rangle\langle\Delta x_2^2\rangle\ge1/4$, where $\Delta$ denotes the difference between an operator and its expectation value, $\Delta x\equiv x-\langle x\rangle$, and hence $\langle\Delta x^2\rangle$ is the variance of $x$.

We can think of the signal as being carried by a single-mode field
\begin{equation}
E(t)=\frac{1}{2}(ae^{-i\omega t}+a^\dagger e^{-i\omega t})
=\frac{1}{\sqrt2}(x_1\cos\omega t+x_2\sin\omega t)\;.
\end{equation}
The annihilation operator is a complex-amplitude operator for the field; the expectation value of the field, $\langle E(t)\rangle=\mbox{Re}(\langle a\rangle e^{-i\omega t})$, oscillates with the amplitude and phase of $\langle a\rangle$.  The variance of $E$ characterizes the noise in the signal; for phase-insensitive noise, this variance is constant and given by $\langle\Delta E^2\rangle=\frac{1}{2}\langle|\Delta a|^2\rangle$, where
\begin{equation}
\langle|\Delta a|^2\rangle
=\frac{1}{2}\bigl\langle\Delta a\Delta a^\dagger+\Delta a^\dagger\Delta a\bigr\rangle
=\langle|\,a\,|^2\rangle-|\langle a\rangle|^2
\label{eq:asecmom}
\end{equation}
is the symmetric variance of~$a$.  Here we use the notation $|\,a\,|^2=\frac{1}{2}(aa^\dagger+a^\dagger a)$ for the symmetric product of $a$ and $a^\dagger$~\cite{Cavesnotes}.

The symmetric variance~(\ref{eq:asecmom}) obeys an uncertainty principle,
\begin{equation}
\langle|\Delta a|^2\rangle
=\frac{1}{2}\!\left(\langle\Delta x_1^2\rangle+\langle\Delta x_2^2\rangle\right)
\ge\langle\Delta x_1^2\rangle^{1/2}\langle\Delta x_2^2\rangle^{1/2}\ge\frac{1}{2}\;.
\label{eq:aup}
\end{equation}
The lower bound is the half-quantum of zero-point (or vacuum) noise.   The first inequality is saturated if and only if the noise is phase insensitive, i.e., $\langle|\Delta a|^2\rangle=\langle\Delta x_1^2\rangle=\langle\Delta x_2^2\rangle$, the second if and only if the quadrature uncertainties have minimum uncertainty product.  Both inequalities are saturated if and only if the mode is in a coherent state $|\alpha\rangle=D(a,\alpha)|0\rangle$, where
\begin{equation}
D(a,\alpha)=e^{\alpha a^\dagger-\alpha^* a}=e^{i(\alpha_2x_1-\alpha_1x_2)}\;,\quad
\alpha=\frac{1}{\sqrt2}(\alpha_1+i\alpha_2)\;,
\end{equation}
is the displacement operator for mode~$a$.  We use a two-slot notation for the displacement operator, partly so as to identify the mode the displacement operator pertains to and partly so that by putting a c-number in both slots, as in $D(\beta,\alpha)=e^{\alpha\beta^*-\alpha^*\beta}=e^{i(\alpha_2\beta_1-\alpha_1\beta_2)}$, we have a convenient notation for two-dimensional Fourier transforms~\cite{Cavesnotes}.

The objective of phase-preserving linear amplification is to increase the size of an input signal by a (real) multiplicative amplitude gain $g$, regardless of the input phase, while introducing as little noise as possible.  The amplification of the input signal can be expressed as the transformation
\begin{equation}
\langle a_{\rm out}\rangle=g\langle a_{\rm in}\rangle
\label{eq:expio}
\end{equation}
of the expected complex amplitude.  A \textit{perfect\/} linear amplifier would perform this feat while preserving the signal-to-noise; in the Heisenberg picture, the primary mode's annihilation operator, not just its expectation value, would transform from input to output~as
\begin{equation}
a_{\rm out}=ga_{\rm in}\;.
\label{eq:ioperfect}
\end{equation}
The second-moment noise would be amplified by the power gain $g^2$, i.e.,
$\langle|\Delta a_{\rm out}|^2\rangle=g^2\langle|\Delta a_{\rm in}|^2\rangle$.  The amplifier's output would be contaminated by the same noise as the input, blown up by a factor of $g^2$, but  the amplification process would not add any noise to the amplified input noise.

Alas, quantum mechanics prohibits free lunches: there are no perfect phase-preserving linear amplifiers; the transformation~(\ref{eq:ioperfect}) does not preserve the canonical commutation relation and thus violates unitarity.  Physically, this is the statement that amplification of the primary mode requires it to be coupled to other physical systems, not least to provide the energy needed for amplification; these other systems, which can thought of as the amplifier's internal degrees of freedom, necessarily add noise to the output.  This physical requirement is expressed in an input-output relation~\cite{Haus1962a,Caves1982a},
\begin{equation}
a_{\rm out}=ga_{\rm in}+L^\dagger\;,
\label{eq:io}
\end{equation}
where the \textit{added-noise operator\/} $L$ is a property of the internal degrees of freedom.  One usually assumes that $\langle L^\dagger\rangle=0$ so as to retain the expectation-value transformation~(\ref{eq:expio}).  Preserving the canonical commutation relation between input and output requires that
\begin{equation}
[L,L^\dagger]=g^2-1\;,
\label{eq:Lcomm}
\end{equation}
which implies an uncertainty principle,
\begin{equation}
\langle|\Delta L|^2\rangle\ge\frac{1}{2}(g^2-1)\;.
\label{eq:addednoise}
\end{equation}

\begin{figure}[htbp]
{\ifnum\figstyle=0 \includegraphics[width=\textwidth]{fig1.eps} \else\includegraphics[width=\textwidth]{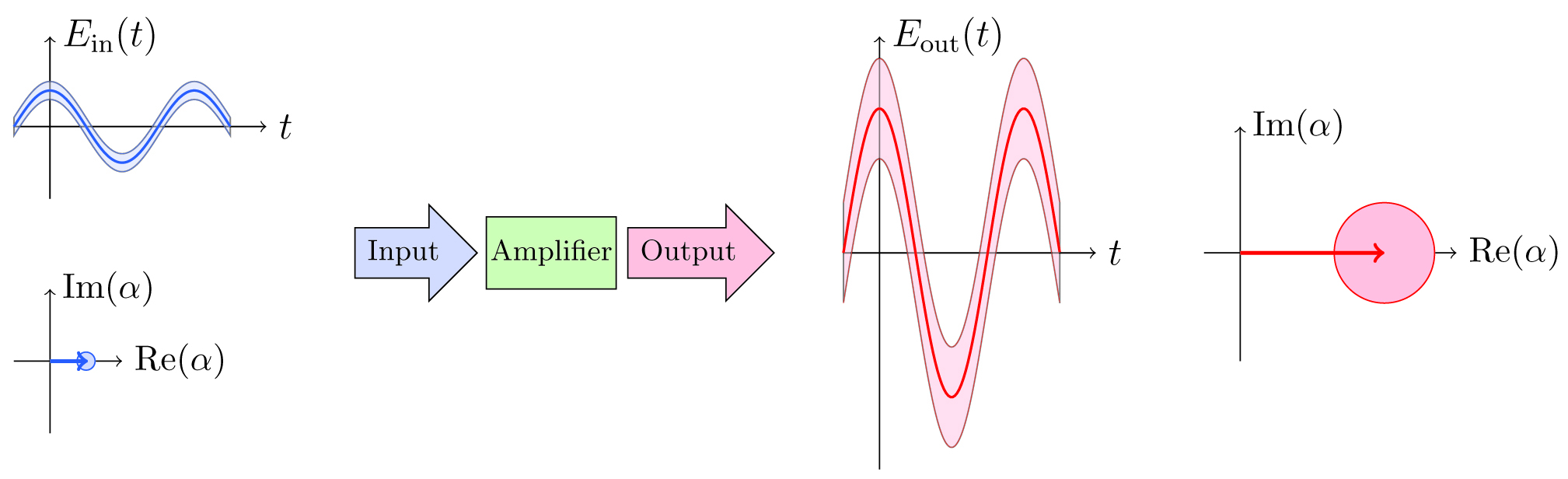} \fi}
\caption{Input signal $E_{\rm in}(t)=\cos\omega t$ (blue) and output signal $E_{\rm out}(t)=g\cos\omega t$ (red) for a single-mode, phase-preserving linear amplifier with amplitude gain $g=4$.  Input and output have the same phase.  The input complex amplitude, $\langle a_{\rm in}\rangle=\langle x_1\rangle/\sqrt2=1$, is amplified to become the output complex amplitude, $\langle a_{\rm out}\rangle=g$.  The (phase-insensitive) noise on the signal is represented by smearing out the mean signal into a band with vertical height equal to the uncertainty in the field, $\langle\Delta E^2\rangle^{1/2}=\langle|\Delta a|^2\rangle^{1/2}/\sqrt2$; for the ideal linear amplification with coherent-state input shown here, the height of the input band is $\frac{1}{2}$, and the height of the output band is $\sqrt{(g^2-\frac{1}{2})/2}$.  The phase-space diagrams depict the same input and output as do the temporal plots.  The mean complex amplitude is represented by an arrow, and the noise by a circle whose diameter is equal to the height of the band in the temporal plot.  The temporal plots can be obtained by rotating the phase-space stick and ball about the origin and projecting onto the real axis.}\label{fig0}
\end{figure}

The amplifier should be prepared to receive any input in the primary mode, without having any idea what that input is going to be.  This places the restriction that the primary mode and the internal degrees of freedom \textit{cannot\/} be correlated before amplification.  The total output noise is then the sum of the amplified input noise and the noise added by the internal degrees of freedom:
\begin{equation}
\langle|\Delta a_{\rm out}|^2\rangle
=g^2\langle|\Delta a_{\rm in}|^2\rangle+\langle|\Delta L|^2\rangle\;.
\end{equation}
The added noise is constrained by the uncertainty principle~(\ref{eq:addednoise}), which together with Eq.~(\ref{eq:aup}), places a lower bound on the output noise:
\begin{equation}
\langle|\Delta a_{\rm out}|^2\rangle\ge g^2-\textstyle{\frac{1}{2}}\;.
\label{eq:aoutup}
\end{equation}
There are, of course, states for which the output noise is smaller---indeed, as small as half a quantum---but these require that the primary mode and the amplifier's internal degrees of freedom be correlated at the input.  Figure~\ref{fig0} illustrates the amplification of the field $E$ and introduces the traditional ball-and-stick phase-space diagrams that are used to depict the second-moment noise.

Referred to the input, the output noise takes the form
\begin{equation}
\frac{\langle|\Delta a_{\rm out}|^2\rangle}{g^2}
=\langle|\Delta a_{\rm in}|^2\rangle+\frac{\langle|\Delta L|^2\rangle}{g^2}
=\langle|\Delta a_{\rm in}|^2\rangle+\mathcal{A}\;.
\label{eq:outputtoinput}
\end{equation}
The second-moment added noise, referred to the input, is called the \textit{added-noise number\/} $\mathcal{A}$~\cite{Caves1982a}.  It provides a natural, dimensionless characterization of an amplifier's performance, and it is constrained by quantum mechanics to satisfy
\begin{equation}
{\cal A}\equiv\frac{\langle|\Delta L|^2\rangle}{g^2}\ge\frac{1}{2}\!\left(1-\frac{1}{g^2}\right)
\mathop{\longrightarrow}_{g\rightarrow\infty}\frac{1}{2}\;.
\end{equation}

Amplifier performance is often characterized by noise temperature~$T_n$ or noise figure $F$.
The noise temperature is defined as the temperature required to account for all of the output noise referred to the input, i.e., to account for the noise~(\ref{eq:outputtoinput}); when the input is quantum limited, i.e., a coherent state with $\langle|\Delta a_{\rm in}|^2\rangle=\frac{1}{2}$, the noise temperature is related to the added-noise number by $\mathcal{A}=(e^{\hbar\omega/k_BT_n}-1)^{-1}$.  The noise figure is the ratio of the input signal-to-noise ratio to the output signal-to-noise ratio; for quantum-limited input, the noise figure is given by $F=1+2\mathcal{A}$.  For quantum-limited input, the noise temperature and noise figure satisfy
\begin{align}
\frac{k_BT_n}{\hbar\omega}=\frac{1}{\ln(1+\mathcal{A}^{-1})}
&\ge\frac{1}{\ln[(3-1/g^2)/(1-1/g^2)]}\mathop{\longrightarrow}_{g\rightarrow\infty}\frac{1}{\ln3}\;,\\
\vphantom{\frac{k_BT_n}{\hbar\omega}}
F=1+2\mathcal{A}&\ge2-1/g^2\mathop{\longrightarrow}_{g\rightarrow\infty}2\;;
\end{align}
an amplifier that operates far from the quantum limit has $\mathcal{A}=F/2=k_BT_n/\hbar\omega\gg1$. In the limit of high gain, a phase-preserving linear amplifier adds at least half a quantum of noise to the input noise, and as a consequence, for a quantum-limited input, the signal-to-noise ratio is degraded by at least a factor of two.

All three of these measures of amplifier performance are afflicted by the residual gain dependence $1-1/g^2$.  The noise temperature and noise figure have the additional annoyance that, as they depend on the input noise, they are not solely properties of the amplifier.  Finally, the noise temperature is not even linear in the added noise for amplifiers operating near the quantum limit.  We prefer in this paper to deal with an added-noise number that has all the gain dependence removed:
\begin{equation}
A_1\equiv\frac{\langle|\Delta L|^2\rangle}{g^2-1}
=\frac{\mathcal{A}}{1-1/g^2}\ge\frac{1}{2}\;.
\label{eq:addednoiseno}
\end{equation}
The second-moment quantum limit is $A_1\ge1/2$.  The subscript indicates that $A_1$ is the first in a sequence of added-noise numbers.  We introduce added-noise numbers for all moments of the added noise in Sec.~\ref{sec:momlim} and consider the limits imposed by quantum mechanics on moments of all orders.

\subsection{Ideal linear amplifier}
\label{subsec:ideal}

An ideal linear amplifier saturates the second-moment bound~(\ref{eq:addednoiseno}).  The added noise in this case is necessarily Gaussian, as we show in Sec.~\ref{sec:genlim}; there are no constraints on an ideal linear amplifier beyond this second-moment bound.  It is instructive to introduce here a pictorial representation of the amplified input noise and the added noise for the case of an ideal linear amplifier acting on a quantum-limited (coherent-state) input.  This allows us to discuss the several perspectives provided by the various ways of ordering creation and annihilation operators.  Since the noise is Gaussian, the pictorial representation can be simplified to ball-and-stick figures, like those in Fig.~\ref{fig0}, which depict only the first and second moments.

\begin{figure}[htbp]
{\ifnum\figstyle=0 \includegraphics[width=0.7\textwidth]{fig2.eps} \else\includegraphics[width=0.7\textwidth]{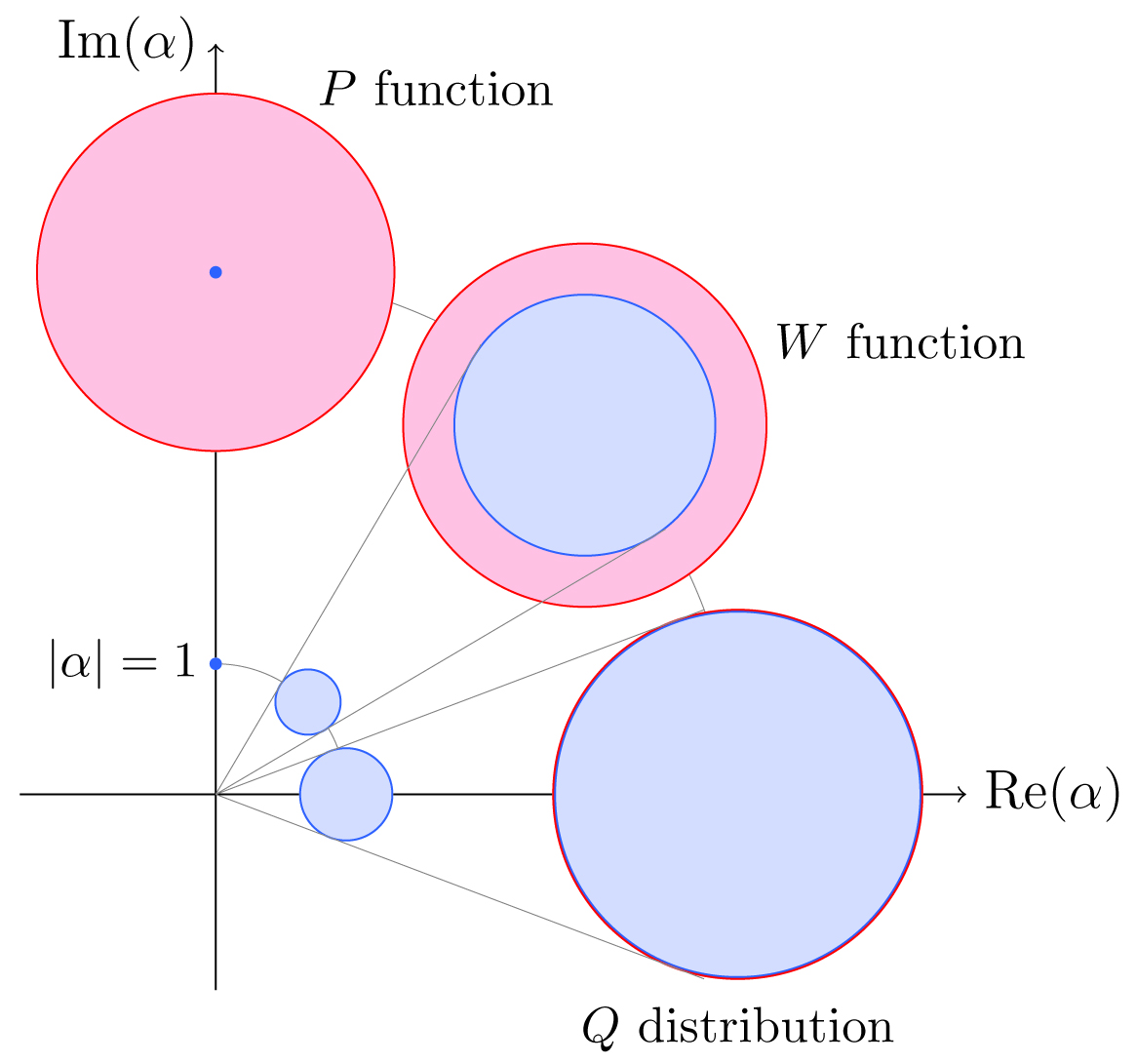} \fi}
\caption{Ball-and-stick phase-space depictions of input and output states for an ideal linear amplifier whose input is a coherent state $|\beta\rangle$ with $|\beta|=1$ and which has amplitude gain $g=4$, giving the output state a mean that lies on a circle of radius $g|\alpha|=4$ centered at the origin.  The three cases correspond to normal ordering (Glauber-Sudarshan $P$ function), symmetric ordering (Wigner $W$ function), and antinormal ordering (Husimi $Q$ distribution).  The input (blue) and output (red) noise are characterized by the variance $\Sigma^2$ of $\alpha$: the radius of the noise circle is chosen to be $\Sigma/2\sqrt2$.  For normal ordering, $\Sigma_P^2=\langle\Delta a^\dagger\Delta a\rangle$; for symmetric ordering, $\Sigma_W^2=\langle|\Delta a|^2\rangle=\frac{1}{2}\langle\Delta a\Delta a^\dagger+\Delta a^\dagger\Delta a\rangle$ ($\Sigma_W/2\sqrt2=\frac{1}{2}\langle\Delta E^2\rangle^{1/2}$); and for antinormal ordering, $\Sigma_Q^2=\langle\Delta a\Delta a^\dagger\rangle$.  At the input, $\Sigma_P^2=0$, $\Sigma_W^2=\frac{1}{2}$, and $\Sigma_Q^2=1$; at the output, $\Sigma_P^2=g^2-1$, $\Sigma_W^2=g^2-\frac{1}{2}$, and $\Sigma_Q^2=g^2$.  For each ordering, blue is used for the input state and for the amplification of the input mean and noise by a factor of $g$; red depicts the total noise at the output.  For symmetric ordering, the output noise consists of the amplified input noise, with variance $\frac{1}{2}g^2$, and added noise, with variance $\frac{1}{2}(g^2-1)$, which degrades the output signal-to-noise ratio relative to the perfect amplification of Eq.~(\protect\ref{eq:ioperfect}).  For normal ordering, the input is represented by a (noiseless) point, which is amplified to another point; all of the output noise, which has variance $g^2-1$, appears to be added noise.  The flip side is antinormal ordering, where the input has an extra half-quantum of noise relative to symmetric ordering; the amplified input noise, which has variance $g^2$, accounts for all of the output noise, and there appears to be no added noise.  Even for the modest gain used here, it is difficult to distinguish the sizes of the output noise circles for the three orderings, because the output noise dwarfs the half-quantum difference between the orderings.}\label{fig1}
\end{figure}

A Gaussian phase-space distribution that has phase-insensitive noise has the form $e^{-|\alpha-\beta|^2/\Sigma^2}/\pi\Sigma^2$, where $\beta$ is the mean value of $\alpha$ and $\Sigma^2$ is the variance of $\alpha$; this variance can be calculated using several orderings, which go with different quasidistributions for the field mode~\cite{Cahill1969a,Cahill1969b,Hillery1984a,GarrisonChiao2008}.  Up till now, we have used the symmetrically ordered variance $\Sigma_W^2=\langle|\Delta a|^2\rangle=2\langle\Delta E^2\rangle$, which goes with the symmetric ordering of the Wigner function \cite{Wigner1932a} and which we thus denote with a subscript $W$.  If we use normal ordering of $a$ and $a^\dagger$, the appropriate quasidistribution is the Glauber-Sudarshan $P$ function~\cite{Glauber1963a,Sudarshan1963a,Glauber1963b}, and the normally-ordered variance is $\Sigma_P^2=\langle\Delta a^\dagger\Delta a\rangle=\Sigma_W^2-\frac{1}{2}$.  Likewise, if we use antinormal ordering, the appropriate quasidistribution is the Husimi $Q$ distribution~\cite{Husimi1940a}, and the antinormally-ordered variance is $\Sigma_Q^2=\langle\Delta a\Delta a^\dagger\rangle=\Sigma_W^2+\frac{1}{2}$.

For an ideal linear amplifier with coherent-state input, we depict the input and output in Fig.~\ref{fig1}, for normal, symmetric, and antinormal ordering of the noise variance.  We use  circles centered at the mean value, with the radius $\Sigma/2\sqrt2$ representing the size of the noise; the stick of Fig.~\ref{fig0} is omitted as redundant.  The multiple of $\Sigma$ we use for the radius of the noise circle is chosen so that the noise circles fit within the figure without overlapping.  One other ingredient appears in Fig.~\ref{fig1}: the amplified input noise is obtained by expanding the input-noise circle by a factor of $g$.  For the case of symmetric ordering, this amplified input noise shows what the output noise would be for a perfect linear amplifier, as in Eq.~(\ref{eq:ioperfect}).  The other two orderings give different perspectives on the amplifier noise, discussed in the figure caption.

Of particular interest is the Husimi $Q$ distribution, which for a modal state $\rho$ is given by $Q_\rho(\alpha)=\langle\alpha|\rho|\alpha\rangle/\pi$.  The stick-figure depiction of Fig.~\ref{fig1} shows that from the perspective of the antinormally-ordered variance, all the output noise in an ideal linear amplifier is amplified input noise, with no added noise at all.  Indeed, Fig.~\ref{fig1} suggests that the output $Q$ distribution of an ideal linear amplifier is a scaled version of the input $Q$, scaled by the gain $g$, i.e., $Q_{\rm out}(\beta)=Q_{\rm in}(\beta/g)/g^2$, and this turns out to be true for arbitrary input states, as we show shortly.  There is a reason why there is no added noise in the $Q$-distribution picture: the nonnegative $Q$ distribution describes the statistics of a quantum-limited, simultaneous measurement of both quadrature components, $x_1$ and $x_2$; the scaling of the $Q$ distribution from input to output says simply that relative to quantum-limited measurements of the signal at both the input and the output, there is no degradation of signal-to-noise ratio between the input and the output of an ideal linear amplifier.  We stress this conclusion: \textit{relative to the best measurements one can make, there is no loss of signal-to-noise in amplifying a signal}.  The reduction in signal-to-noise occurs in symmetrically ordered moments, not in the antinormally ordered moments that apply to simultaneous measurements of both quadrature components.

The goal of this paper is to go beyond the Gaussian noise of an ideal linear amplifier.  Thus we will need to move beyond the ball-and-stick diagrams of Fig.~\ref{fig1} and plot the entire distributions of input and output noise.  In doing so, we will make use of all three operator orderings, moving freely among them as we find one or the other better serves our purpose.  We find it most convenient to formulate our general mathematical description of a phase-preserving linear amplifier in terms of the $P$-function picture, though we quickly generalize to all three orderings.   Before turning to that task, which occupies Sec.~\ref{sec:char}, we review several models of an ideal linear amplifier.

We also note complementary work on the quantum noise limits for operational amplifiers~\cite{Courty1999a,Clerk2004a,Clerk2010a}.  An operational amplifier takes in the voltage, say, at the end of a transmission line and outputs a voltage into another transmission line.  Since the input and output voltages both consist of incident and reflected waves, an op-amp does not have the simple input-output structure that is the basis of our work here and most previous work on linear amplifiers.

\subsection{Models for ideal linear amplifiers}\label{subsec:idealmodels}

In this subsection we consider models for ideal linear amplifiers.  While detailed microscopic models of particular amplifiers and their noise sources (see, e.g.,~\cite{CarWal74,ZShi2011a}) are important, they are not central to our analysis.  Here we survey four generic models of a linear amplifier, to build intuition about the fundamental physical processes that account for amplifier noise and to provide context for our subsequent study of general quantum constraints on the performance of linear~amplifiers.

\subsubsection{Parametric amplifier}

The simplest model of an ideal linear amplifier is provided by a parametric amplifier~\cite{Mollow1967a,Mollow1967b,Collett1988a}.  The primary mode $a$ interacts with an ancillary mode $b=(y_1+iy_2)/\sqrt2$, which is initially in the vacuum state.  The total Hamiltonian,
\begin{equation}
H=\hbar\omega(a^\dagger a+b^\dagger b)
+i\hbar\kappa\bigl(abe^{2i\omega t}-a^\dagger b^\dagger e^{-2i\omega t}\bigr)\;,
\end{equation}
has an interaction term that describes pairwise creation or destruction of quanta in the two modes. This pairwise creation or destruction is accompanied by destruction or creation of a quantum in a pump mode that has frequency $2\omega$.  The pump mode does not appear in the Hamiltonian because it is excited into a high-amplitude coherent state and thus is essentially classical.  Its amplitude contributes to the coupling strength $\kappa$, and its time dependence gives the $e^{\pm2i\omega t}$ explicit time dependences in the Hamiltonian.

Transforming to the interaction picture that removes the free Hamiltonians of the two modes, the interaction part of the Hamiltonian assumes the form
\begin{equation}
H_I=i\hbar\kappa(ab-a^\dagger b^\dagger)\;,
\label{eq:parampHI}
\end{equation}
which can be integrated to give an evolution operator
\begin{equation}
U_I(t)=e^{-iH_It/\hbar}=\exp\bigl[r(ab-a^\dagger b^\dagger)\bigr]=
\exp\bigl[ir(x_1y_2+x_2y_1)\bigr]\equiv S(r)\;,
\quad\mbox{$r=\kappa t$\;,}
\label{eq:Sr}
\end{equation}
where $S(r)$ is the \textit{two-mode squeeze operator\/}~\cite{Caves1985a,Schumaker1985a,Schumaker1986a}.
In the Heisenberg picture, the primary mode's annihilation operator undergoes the transformation
\begin{equation}
a_{\rm out}=S^\dagger aS=a\cosh r-b^\dagger\sinh r=ga-b^\dagger\sqrt{g^2-1}\;,
\label{eq:iotwo}
\end{equation}
i.e., the amplifier input-output relation~(\ref{eq:io}) with gain $g=\cosh r$ and noise operator $L=-b\sinh r=-b\sqrt{g^2-1}$.  If the ancillary mode begins in the vacuum state $|0\rangle$, the noise operator saturates the second-moment bound~(\ref{eq:addednoiseno}), and we have an ideal linear amplifier.

\begin{figure}[htbp]
{\ifnum\figstyle=0 \includegraphics[width=0.8\textwidth]{fig3.eps} \else\includegraphics[width=0.8\textwidth]{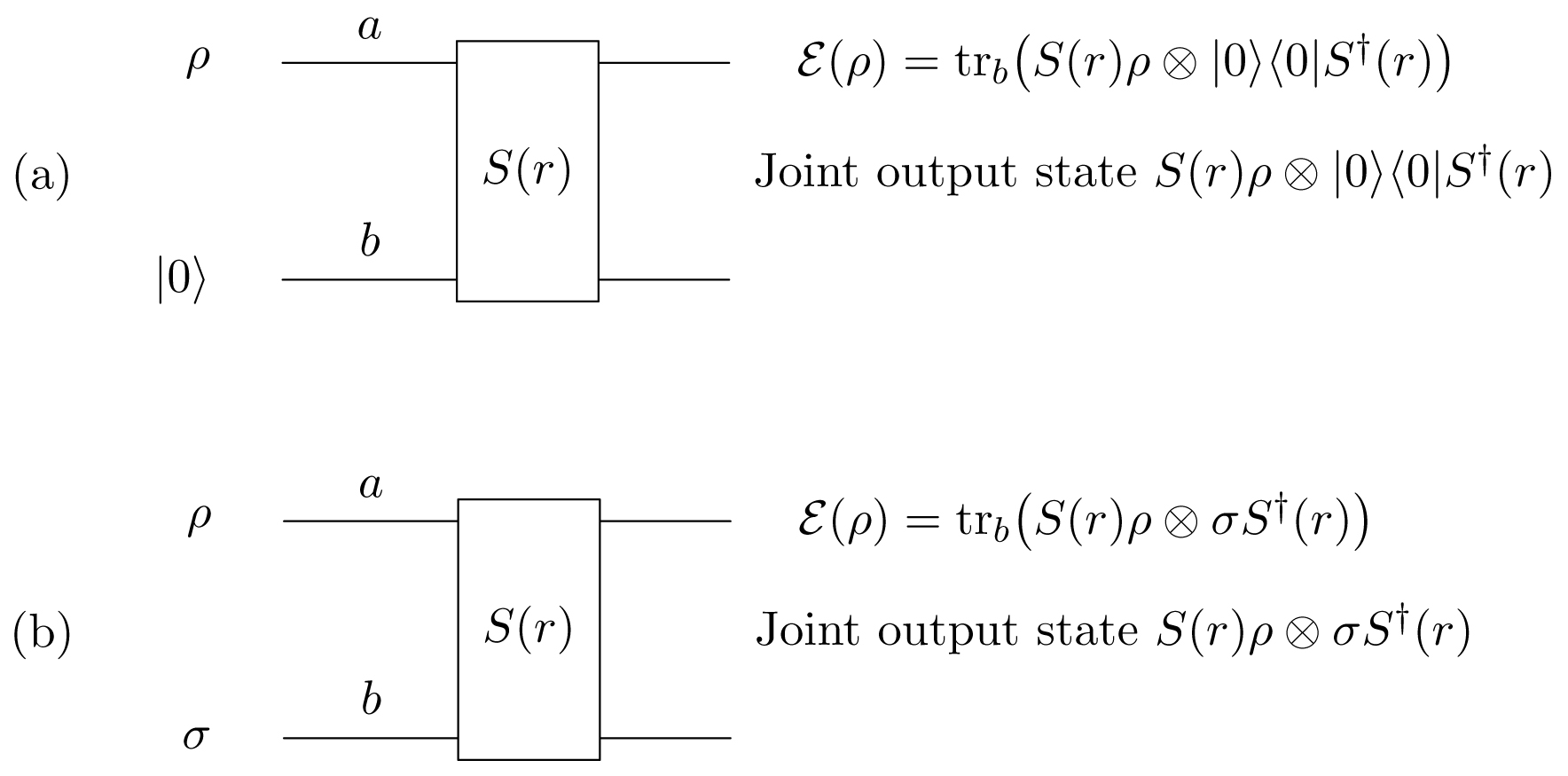} \fi}
\caption{(a)~Ideal phase-preserving linear amplifier realized as a parametric amplifier.  The primary mode~$a$ interacts with an ancillary mode~$b$, which begins in the vacuum state $|0\rangle$, via a two-mode squeezing interaction.  The output state $\sE(\rho)$ of the primary mode is amplified with an amplitude gain $g=\cosh r$ and with the minimum amount of added noise permitted by quantum mechanics.  (b)~Parametric amplifier with an arbitrary initial state $\sigma$ for the ancillary mode.  The main result of this paper is that \textit{any\/} phase-preserving linear amplifier, no matter its physical realization, is equivalent to a parametric amplifier with some physical initial state $\sigma$ for the ancillary mode.}
\label{fig2}
\end{figure}

We can also describe the evolution in the interaction picture.  If the primary mode has initial state $\rho$ and the ancillary mode begins in vacuum, the state of the primary mode after amplification~is
\begin{equation}
\sE(\rho)=\tr_b\bigl(S(r)\rho\otimes|0\rangle\langle0|S^\dagger(r)\bigr)\;.
\label{eq:Eideal}
\end{equation}
Here $\sE$ is the trace-preserving quantum operation (a completely positive map) that describes how the state of primary mode transforms from the input to the output of the amplifier.   Figure~\ref{fig2}(a) gives the simple quantum circuit for an ideal parametric amplifier.

The factored expression for the squeeze operator~\cite{Schumaker1985a,Schumaker1986a},
\begin{equation}
S(r)=\frac{1}{g}e^{-a^\dagger b^\dagger\tanh r}g^{-(a^\dagger a+b^\dagger b)}e^{ab\tanh r}
=\frac{1}{g}g^{-a^\dagger a}e^{-a^\dagger b^\dagger\sinh r}e^{ab\sinh r}g^{-b^\dagger b}\;,
\label{eq:Sfactored}
\end{equation}
can be used to eliminate the ancillary mode from the ideal-amplifier quantum operation~(\ref{eq:Eideal}).  In a first approach, we find the partial matrix element of $S(r)$ between a coherent state $|\beta\rangle$ and vacuum for the ancillary mode:
\begin{equation}
\langle\beta|S(r)|0\rangle
=\frac{e^{-|\beta|^2/2}}{g}g^{-a^\dagger a}e^{-\sqrt{g^2-1}\beta^* a^\dagger}
\equiv \sqrt{\pi}A_\beta\;.
\end{equation}
Taking the trace in Eq.~(\ref{eq:Eideal}) in the coherent-state basis of the ancillary mode,
\begin{equation}
\sE(\rho)=\int\frac{\dtwo\beta}{\pi}\,\langle\beta|S|0\rangle\rho\langle0|S^\dagger|\beta\rangle
=\int\dtwo\beta\,A_\beta\rho A_\beta^\dagger
\end{equation}
($\dtwo\beta=d\beta_R\,d\beta_I=\frac{1}{2}d\beta_1d\beta_2$), gives a Kraus decomposition of $\sE$; the operators $A_\beta$ are called Kraus operators~\cite{Kraus1983a,Nielsen2000a}.  One can use this to find the output $Q$ distribution
for an ideal linear amplifier.  As promised above, the output $Q$ distribution is a scaled version of the input $Q$ distribution:
\begin{equation}
Q_{\rm out}(\alpha)
=\frac{1}{\pi}\bigl\langle\alpha\bigl|\sE(\rho)\bigr|\alpha\bigr\rangle
=\frac{Q_{\rm in}(\alpha/g)}{g^2}\;.
\label{eq:Qio}
\end{equation}

A different Kraus decomposition of $\sE$~\cite{Nha2010a} follows from partial matrix elements of the squeeze operator in the number basis of mode~$b$:
\begin{equation}
\langle n|S(r)|0\rangle=
\frac{(-1)^n}{g}\frac{(g^2-1)^{n/2}}{\sqrt{n!}}g^{-a^\dagger a}(a^\dagger)^n\equiv A_n\;.
\end{equation}
Taking the trace in Eq.~(\ref{eq:Eideal}) in the number basis of the ancillary mode gives the Kraus decomposition
\begin{equation}
\sE(\rho)=\sum_{n=0}^\infty\langle n|S|0\rangle\rho\langle0|S^\dagger|n\rangle
=\sum_{n=0}^\infty A_n\rho A_n^\dagger\;.
\end{equation}

\subsubsection{Inverted-oscillator model}

Another model for an ideal linear amplifier, due to Glauber~\cite{Glauber1986a}, uses a primary mode~$a$ and an ancillary mode~$b$.  The two modes have the same frequency $\omega$, but the ancillary mode is an inverted oscillator (sometimes called a negative-mass oscillator~\cite{MTsang2010c,MTsang2011a,MTsang2012au}).  The inverted oscillator has an upside-down Hamiltonian, $-\hbar\omega b^\dagger b$.  Since its energy levels run down instead of up, the inverted oscillator is a source of energy; when a quantum is created in the inverted oscillator, the oscillator emits energy $\hbar\omega$.  The Hamiltonian of the two modes is
\begin{equation}
H=\hbar\omega(a^\dagger a-b^\dagger b)+i\hbar\kappa(ab-a^\dagger b^\dagger)\;.
\end{equation}
Transforming to an interaction picture that removes the free Hamiltonians gives the interaction Hamiltonian~(\ref{eq:parampHI}) of a parametric amplifier; the subsequent discussion of linear amplification is thus identical to that for a parametric amplifier.  Indeed, the only difference between a parametric amplifier and the inverted-oscillator model is that the inverted oscillator does not need a pump to balance the energy books; creation of a quantum in the inverted oscillator provides the energy needed to create a quantum in the primary mode.

\subsubsection{Linear-amplifier master equation}

Our third model is an elaboration of either a parametric amplifier or the inverted-oscillator model.  The single ancillary mode is replaced by a field, which is initially in the vacuum state.  The instantaneous temporal field modes interact with the primary mode via a parametric interaction like Eq.~(\ref{eq:parampHI}); the result is the master equation for an ideal linear amplifier~\cite{CGardiner2004a}.  Models of this sort, based on   coupling the primary mode to a sequence of inverted oscillators, have been developed by several authors~\cite{Jeffers1993a,Loudon2000a,CGardiner2004a}.

This approach starts with an (interaction) Hamiltonian
\begin{equation}
H=i\hbar\sqrt\gamma\int_0^\infty d\tau\,(a b_\tau-a^\dagger b^\dagger_\tau)\delta(\tau-t)
=i\hbar\sqrt\gamma(ab_t-a^\dagger b_t^\dagger)\;.
\end{equation}
The operators $b_\tau$ and $b_\tau^\dagger$ are continuum annihilation and creation operators for the instantaneous field modes, obeying the canonical commutator $[b_\tau,b_{\tau'}^\dagger]=\delta(\tau-\tau')$.  The parameter $\tau$ labels the field modes and specifies the time $t=\tau$ at which a field mode interacts with the primary mode.

It is easy to derive the Heisenberg-picture equations of motion:
\begin{align}
\label{eq:meadot}
\frac{da}{dt}&=\frac{1}{i\hbar}[a,H]=-\sqrt\gamma b_t^\dagger(t)\;,\\
\frac{db_\tau}{dt}&=\frac{1}{i\hbar}[b_\tau,H]=-\sqrt\gamma a^\dagger(t)\delta(t-\tau)\;.
\label{eq:mebdot}
\end{align}
The solution of Eq.~(\ref{eq:mebdot}),
\begin{equation}
b_\tau(t)=b_\tau(0)-\sqrt\gamma a^\dagger(\tau)\Theta(t-\tau)\;,
\end{equation}
where $\Theta(t)$ is the unit step function, with value $\frac{1}{2}$ at $t=0$, can be plugged into Eq.~(\ref{eq:meadot}) to give
\begin{equation}
\frac{da}{dt}=\frac{1}{2}\gamma a(t)-\sqrt\gamma b_t^\dagger(0)\;,
\end{equation}
whose solution is
\begin{equation}
a(t)=g(t)a(0)+L^\dagger(t)\;.
\end{equation}
Here
\begin{equation}
g(t)=e^{\gamma t/2}
\label{eq:gt}
\end{equation}
is the amplitude gain that applies if the interaction is turned off at time $t$, and
\begin{equation}
L(t)=-\sqrt\gamma\int_0^t d\tau\,g(t-\tau)b_\tau(0)
\end{equation}
is the added-noise operator.  It is easy to verify that $L(t)$ and $L^\dagger(t)$ satisfy the commutator~(\ref{eq:Lcomm}), as required by unitarity.  One can think of all the added noise as coming from a single, discrete, wave-packet mode, whose annihilation operator is $b=L/\sqrt{g^2-1}$.

The easiest way to derive the corresponding master equation is to discretize the field modes into wave packets, each of which lasts a short time $\Delta t$:
\begin{equation}
b_j=\frac{1}{\sqrt{\Delta t}}\int_{t_{j-1}}^{t_j}d\tau\,b_\tau\;.
\end{equation}
Here $t_j=j\Delta t$.  The primary mode interacts sequentially with these discretized modes, according to the interaction Hamiltonian~(\ref{eq:parampHI}) with coupling constant $\kappa=\sqrt{\gamma/\Delta t}$.  The interaction of the primary mode with the $j$th discrete mode
changes the state of the primary mode according to
\begin{equation}
\rho(t_j)=\tr_{b_j}\bigl(S(\kappa\Delta t)
\rho(t_{j-1})\otimes|0\rangle\langle0|S^\dagger(\kappa\Delta t)\bigr)\;.
\end{equation}
Expanding the squeeze operators to second order and rewriting in terms of the change in $\rho$
through the $j$th interaction gives
\begin{equation}
\frac{\Delta\rho}{\Delta t}=\frac{1}{2}{\kappa^2\Delta t}
\bigl(2a^\dagger\rho a-aa^\dagger\rho-\rho aa^\dagger\bigr)\;.
\end{equation}
We now take the limit $\Delta t\rightarrow0$ and $\kappa\rightarrow\infty$, with $\kappa^2\Delta t=\gamma$ held constant, obtaining
\begin{equation}
\frac{d\rho}{dt}=\frac{\gamma}{2}
\bigl(2a^\dagger\rho a-aa^\dagger\rho-\rho aa^\dagger\bigr)\;.
\label{eq:mastereq}
\end{equation}
This is the (ideal) linear-amplifier master equation~\cite{CGardiner2004a}.

Solving the master equation is easy.  Using the standard rules~\cite{Cahill1969a}, translate Eq.~(\ref{eq:mastereq}) to a partial differential equation for the $Q$ distribution:
\begin{equation}
0=\left[\frac{\partial}{\partial t}
+\frac{\gamma}{2}\left(\frac{\partial}{\partial\alpha}\alpha+\frac{\partial}{\partial\alpha^*}\alpha^*\right)\right]
Q(\alpha,\alpha^*;t)\;.
\end{equation}
The solution,
\begin{equation}
Q(\alpha,\alpha^*;t)=\frac{1}{g^2(t)}Q\biggl(\frac{\alpha}{g(t)},\frac{\alpha^*}{g(t)};0\biggr)\;,
\end{equation}
where $g(t)$ is the amplitude gain~(\ref{eq:gt}), shows again that the input-output transformation for an ideal linear amplifier is a rescaling of the input $Q$ distribution by the amplitude gain.

\subsubsection{Measurement-based model of linear amplification}

The last model we consider explores the connection between ideal linear amplification and quantum-limited simultaneous measurements of both quadrature components~\cite{Arthurs1965a}.  The strategy for amplification is to measure both quadrature components of the primary mode, $x_1$ and $x_2$, and then to create an amplified coherent state $|g\alpha\rangle$, where $\alpha=(\alpha_1+i\alpha_2)/\sqrt2$ is determined by the measurement outcomes, $\alpha_1$ and $\alpha_2$.  Discarding the outcomes introduces an average of the output coherent states over the probability distribution for the measurement outcomes.  The result is a linear-amplification process, with the added noise due to the noise that accompanies a simultaneous measurement of $x_1$ and $x_2$.  This is a silly strategy for making a linear amplifier, because the point of linear amplification is to make a small signal accessible without the need for quantum-limited measurements.  Nonetheless, this approach is instructive in highlighting the connection between quantum-limited amplification and quantum-limited measurements of the quadrature components.

\begin{figure}[htbp]
{\ifnum\figstyle=0 \includegraphics[width=\textwidth]{fig4.eps} \else\includegraphics[width=\textwidth]{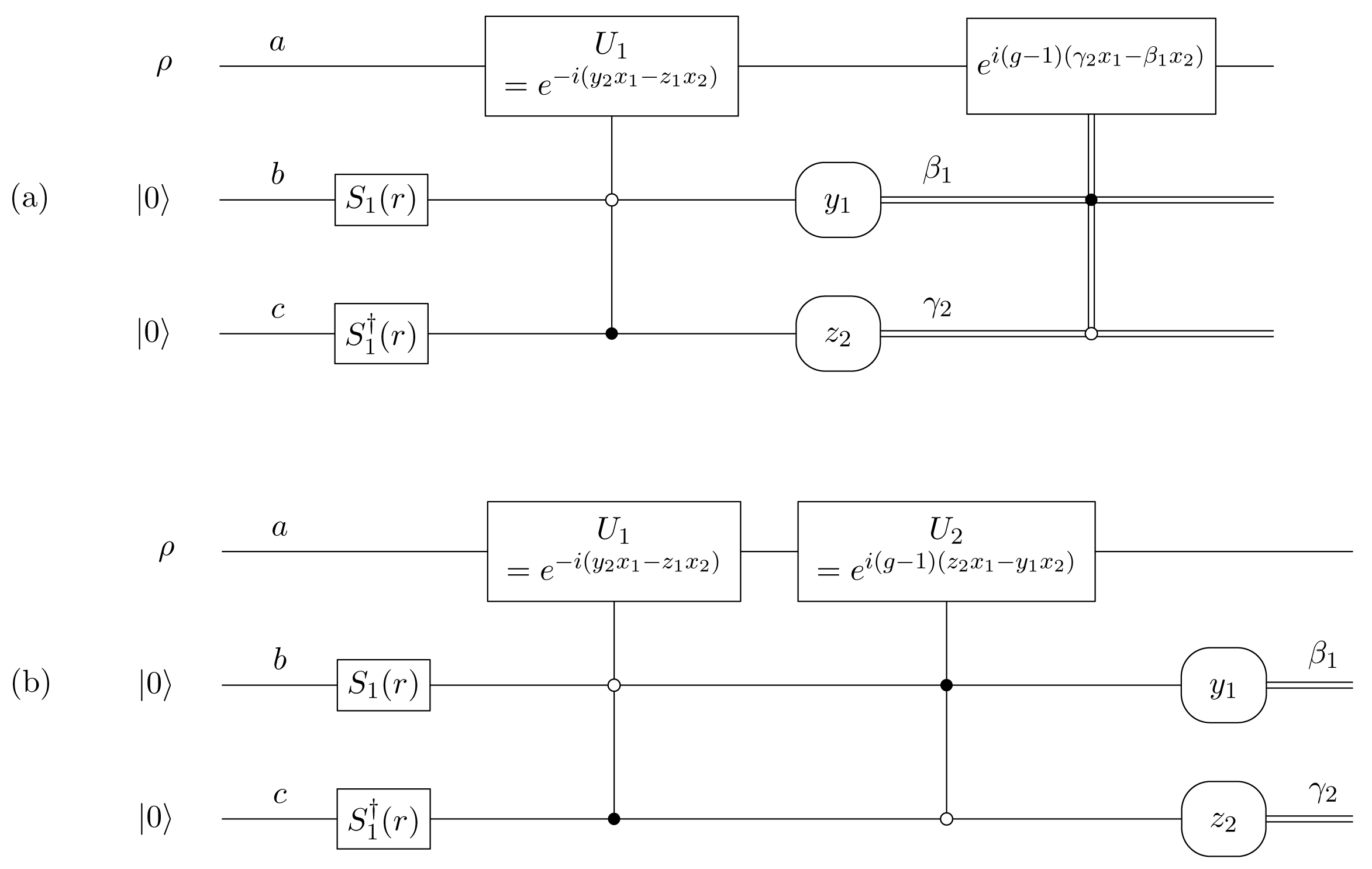} \fi}
\caption{(a)~Quantum circuit for measurement-based amplification of primary mode~$a=(x_1+ix_2)/\sqrt2$. (b)~Circuit for the equivalent coherent scheme.  Filled circles represent control on the first quadrature of a mode; open circles represent control on the second quadrature.  In both cases, the initial controlled operation is a sensing step in which the two quadratures, $x_1$ and $x_2$, of mode~$a$ are written onto ancillary modes $b=(y_1+iy_2)/\sqrt2$ and~$c=(z_1+iz_2)/\sqrt2$; $x_1$ is recorded in $y_1$, and $x_2$ is recorded in $z_2$.  This sensing step is followed by a feedback operation from $b$ and $c$ to amplify the complex amplitude of~$a$.  In (a) the feedback operation is classical and based on the results, $\beta_1$ and $\gamma_2$, of measuring $y_1$ and $z_2$ (measurements are represented by rounded boxes); discarding the measurement outcomes yields a linear-amplification process on the primary mode.  In (b) the feedback operation is coherent; the final measurement is irrelevant to the amplification and can be omitted.  The amplified output state~(\protect{\ref{eq:AKouttwo}}) of mode~$a$ is obtained by tracing out modes $b$ and $c$.  The equivalence of the two circuits, demonstrated in the text, is obvious from the circuit diagrams as an example of the principle of deferred measurement~\protect{\cite{Nielsen2000a,Griffiths1996b}}.  The single-mode squeezers applied initially to the ancillary modes prepare the input state $|\phi\rangle$ of Eq.~(\protect{\ref{eq:phi}}).  When the squeeze parameter~$r$ is chosen to squeeze by a factor of two ($e^{2r}=2$), the model is based on the Arthurs-Kelly simultaneous measurement of $x_1$ and $x_2$~\protect{\cite{Arthurs1965a}}, which is described by coherent-state projectors, but the result is not quite an ideal linear amplifier.  When the squeeze parameter is specified by $e^{2r}=2(g-1)/(g+1)$, the result is an ideal linear amplifier, based on a slightly different model of simultaneous measurement of $x_1$ and $x_2$; this realization of an ideal linear amplifier is different from a parametric amplifier.}
\label{fig3}
\end{figure}

A quantum-limited simultaneous measurement of both quadrature components is described by coherent-state projectors.  The Kraus operators for such a measurement are
\begin{equation}
K_\alpha=\frac{1}{\sqrt\pi}|\alpha\rangle\langle\alpha|\;.
\label{eq:Krauscoh}
\end{equation}
The trace-decreasing quantum operation for outcome $\alpha$ is $K_\alpha\rho K_\alpha^\dagger=Q_\rho(\alpha)|\alpha\rangle\langle\alpha|$; i.e., the measurement statistics are given by the $Q$ distribution, and the output state is the coherent state that corresponds to the measurement outcomes.  The quantum-circuit diagrams in Fig.~\ref{fig3} summarize pictorially the results of the algebraic contortions in the following analysis.

Our goal is to implement the Kraus operators~(\ref{eq:Krauscoh}) in an ancilla model.  To do so, we introduce two ancillary modes, $b=(y_1+iy_2)/\sqrt2$ and $c=(z_1+iz_2)/\sqrt2$, which serve as meters that record the measurement results.  For mode $b$, we introduce $\delta$-normalized eigenstates of the quadrature components, $y_j|\beta_j\rangle=\beta_j|\beta_j\rangle$, $j=1,2$, with $\langle\beta_1|\beta_2\rangle=e^{i\beta_1\beta_2}/\sqrt{2\pi}$; similarly, for $c$, $z_j|\gamma_j\rangle=\gamma_j|\gamma_j\rangle$, $j=1,2$, with $\langle\gamma_1|\gamma_2\rangle=e^{i\gamma_1\gamma_2}/\sqrt{2\pi}$.  In the ancilla model, the result $\alpha_1$ of measuring $x_1$ is recorded in the first quadrature, $y_1$, of mode $b$; thus we identify $\alpha_1$ with $\beta_1$.  Similarly, the result $\alpha_2$  of measuring $x_2$ is recorded in the second quadrature, $z_2$, of mode $c$; thus we identify $\alpha_2$ with $\gamma_2$.

We begin the analysis by using~\cite{Cahill1969a,Cahill1969b,Cavesnotes}
\begin{equation}
|\alpha\rangle\langle\alpha|=
\int\frac{\dtwo\mu}{\pi}\,e^{-\mu^* a}e^{\mu a^\dagger}D(\mu,\alpha)
=\int\frac{\dtwo\mu}{\pi}\,e^{-|\mu|^2/2}D(a,\mu)D(\mu,\alpha)
\label{eq:cohFT}
\end{equation}
to manipulate $K_\alpha$ into the form~\cite{Braunstein1991a}
\begin{equation}
K_\alpha
=\frac{1}{\sqrt\pi}\int\frac{d\beta_2\,d\gamma_1}{2\pi}\,
e^{-(\beta_2^2+\gamma_1^2)/4}e^{-i(\beta_2x_1-\gamma_1x_2)}e^{i(\beta_1\beta_2-\gamma_1\gamma_2)}\;.
\label{eq:Kalpha}
\end{equation}
Here we relabel the measurement outcomes as described above, i.e., $\alpha_1=\beta_1$ and $\alpha_2=\gamma_2$, so $\alpha=(\beta_1+i\gamma_2)/\sqrt2$, and we set $\mu=-(\gamma_1+i\beta_2)/\sqrt2$.  We let the initial state $|\phi\rangle$ of the ancillary modes be specified by the wave function $\langle\beta_2,\gamma_1|\phi\rangle=e^{-(\beta_2^2+\gamma_1^2)/4}/\sqrt{2\pi}$.  This initial state can be written as
\begin{equation}
|\phi\rangle=S_1(r)\otimes S_1^\dagger(r)|0,0\rangle\;,
\label{eq:phi}
\end{equation}
where $|0,0\rangle$ is the vacuum state of modes $b$ and $c$ and
\begin{equation}
S_1(r)=\exp\bigl[\textstyle{\frac{1}{2}}r(b^2-b^{\dagger 2})\bigr]
=\exp\bigl[\textstyle{\frac{1}{2}}ir(y_1y_2+y_2y_1)\bigr]
\label{eq:Soner}
\end{equation}
is the \textit{single-mode squeeze operator\/} for mode $b$ (similarly for mode~$c$)~\cite{Cavesnotes,Caves1985a,Schumaker1985a,Schumaker1986a}, and where the squeeze parameter here corresponds to squeezing by a factor of $2$, i.e., $e^{2r}=2$.  In the state $|\phi\rangle$, the four quadrature components are uncorrelated and have variances
\begin{align}
\langle\Delta y_1^2\rangle
&=\langle\Delta z_2^2\rangle=\frac{1}{2}e^{-2r}=\frac{1}{4}\;,\nonumber\\
\langle\Delta y_2^2\rangle
&=\langle\Delta z_1^2\rangle=\frac{1}{2}e^{2r}=1\;.
\label{eq:varphione}
\end{align}

We can now write the Kraus operator~(\ref{eq:Kalpha}) in the form
\begin{align}
K_\alpha=\sqrt2\int d\beta_2\,d\gamma_1\,\langle\beta_1,\gamma_2|\beta_2,\gamma_1\rangle
\langle\beta_2,\gamma_1|e^{-i(y_2x_1-z_1x_2)}|\phi\rangle
=\sqrt2\langle\beta_1,\gamma_2|U_1|\phi\rangle\;,
\end{align}
where
\begin{equation}
U_1=\exp\bigl[-i(y_2x_1-z_1x_2)\bigr]\;.
\end{equation}
This demonstrates one way to implement $K_\alpha$: let the primary mode interact with the two ancillary modes via an instantaneous interaction $H=\hbar(y_2x_1-z_1x_2)\delta(t)$; then measure $y_1$ and $z_2$ on the ancillary modes, getting results $\beta_1$ and $\gamma_2$, giving a Kraus operator $K_{\beta_1,\gamma_2}=K_\alpha/\sqrt2$.  The $\sqrt2$ comes from the change in integration measure in going from $\alpha=(\beta_1+i\gamma_2)/\sqrt2$ to $\beta_1$ and $\gamma_2$.  The interaction displaces the first $b$-quadrature, $y_1$, by $x_1$ and the second $c$-quadrature, $z_2$, by $x_2$; the measurements of $y_1$ and $z_2$ read out these displacements, contaminated by the uncertainties in $y_1$ and $z_2$, and leave the primary mode in the coherent state~$|\alpha\rangle$ corresponding to the measurement results, $\alpha_1=\beta_1$ and $\alpha_2=\gamma_2$.

To turn this into an amplifier model, we use the result of the measurement to displace the primary mode by $(g-1)\alpha$.  The entire procedure is then described by Kraus operators
\begin{equation}
K'_\alpha=D\bigl(a,(g-1)\alpha\bigr)K_\alpha=\frac{1}{\sqrt\pi}|g\alpha\rangle\langle\alpha|\;.
\end{equation}
Writing this in terms of quadrature components of the ancillary modes gives
\begin{equation}
K'_{\beta_1,\gamma_2}=K'_\alpha/\sqrt2=
e^{i(g-1)(\gamma_2x_1-\beta_1x_2)}\langle\beta_1,\gamma_2|U_1|\phi\rangle
=\langle\beta_1,\gamma_2|U_2U_1|\phi\rangle\;,
\label{eq:Kprime}
\end{equation}
where
\begin{equation}
U_2=\exp\bigl[i(g-1)(z_2x_1-y_1x_2)\bigr]\;.
\end{equation}
In the first form in Eq.~(\ref{eq:Kprime}), there is a sensing interaction $U_1$ and then measurements of $y_1$ on mode~$b$ and $z_2$ on mode~$c$; this is followed by an amplifying displacement $D\bigl(a,(g-1)\alpha\bigr)=e^{i(g-1)(\gamma_2x_1-\beta_1x_2)}$ of mode~$a$ based on the measurement results $\beta_1$ and $\gamma_2$.  Discarding the measurement outcomes leads to a linear amplifier.  In the second form, there are two coherent interactions, first the sensing interaction $U_1$ and then an amplifying feedback $U_2$, and these two are followed by the measurements of $y_1$ and $z_2$; in this second form, discarding the measurement outcomes can be accomplished by omitting the closing measurement.  These considerations are summarized in quantum-circuit diagrams in Fig.~\ref{fig3}.

The output state of this linear amplifier is
\begin{equation}
\sE(\rho)=\int\dtwo\alpha\,K'_\alpha\rho K^{\prime\dagger}_\alpha=
\int\dtwo\alpha\,\frac{Q_\rho(\alpha/g)}{g^2}|\alpha\rangle\langle\alpha|\;.
\label{eq:AKout}
\end{equation}
This can also be written as
\begin{equation}
\sE(\rho)=\tr_{b,c}\bigl(U_2U_1\rho\otimes|\phi\rangle\langle\phi|U_1^\dagger U_2^\dagger\bigr)\;.
\label{eq:AKouttwo}
\end{equation}
It is easy to see from Eq.~(\ref{eq:AKout}) that this is not quite an ideal linear amplifier: the output $P$ function, not the output $Q$ distribution, is a rescaled input $Q$ distribution.  This means that this amplifier adds two more units of vacuum noise than does an ideal linear amplifier.  We can see this more directly---and see also how to convert to ideal linear amplification---by examining the input-output relation for the primary mode's annihilation operator,
\begin{equation}
a_{\rm out}=U_1^\dagger U_2^\dagger aU_2U_1=ga+\sqrt{g^2-1}d^\dagger\;.
\label{eq:AKio}
\end{equation}
Here we introduce a modal annihilation operator $d=(s_1+is_2)/\sqrt2$, where
\begin{align}
s_1&=-\sqrt{\frac{g-1}{g+1}}y_1+\frac{1}{2}\sqrt{\frac{g+1}{g-1}}z_1\;,\\
s_2&=-\frac{1}{2}\sqrt{\frac{g+1}{g-1}}y_2+\sqrt{\frac{g-1}{g+1}}z_2\;.
\end{align}
Since the original quadrature components are uncorrelated, with variances~(\ref{eq:varphione}), $s_1$ and $s_2$ are also uncorrelated, with variances
\begin{equation}
\langle\Delta s_1^2\rangle=\langle\Delta s_2^2\rangle
=\frac{1}{2}\frac{g^2+1}{g^2-1}=\langle|\Delta d|^2\rangle\;.
\end{equation}
The added noise, $\frac{1}{2}(g^2+1)$, is indeed two vacuum units bigger than the added noise~(\ref{eq:addednoise}) of an ideal linear amplifier.

In the high-gain limit, these two additional units of vacuum noise become irrelevant.  On the other hand, it is easy to see how to convert this measurement-based model of linear amplification into an ideal linear amplifier.  What needs to be done is to make the initial state $|\phi\rangle$ of modes $b$ and $c$ the vacuum state of $d$.  Noticing that
\begin{equation}
d=S_1(r)\otimes S_1^\dagger(r)\frac{1}{\sqrt2}(-b+c)S_1^\dagger(r)\otimes S_1(r)\;,
\end{equation}
where the squeeze parameter is specified by
\begin{equation}
e^{2r}=2\,\frac{g-1}{g+1}\;,
\end{equation}
we realize that all we need to do is to use an initial state of the form~(\ref{eq:phi}), with $r$ specified by this new value.  The result is ideal linear amplification by a mechanism that is distinct from a parametric amplifier (see Fig.~\ref{fig3}): the primary mode has an input-output relation~(\ref{eq:AKio}) that is the same as for a parametric amplifier; it is not hard, but tedious to check that the mode~$d$, which is completely responsible for the added noise, does not evolve as does the single ancillary mode of a parametric amplifier.

We can chase this new choice of initial state for the ancillary modes back through the above analysis to see what kind of measurement of $x_1$ and $x_2$ it corresponds to.  The initial wave function for the ancillary modes becomes
\begin{equation}
\langle\beta_2,\gamma_1|\phi\rangle=\frac{1}{\sqrt\pi e^r}\exp\!\left(-\frac{\beta_2^2+\gamma_1^2}{2e^{2r}}\right)\;,
\end{equation}
and this leads ultimately to the following Kraus operators for the simultaneous measurement of $x_1$ and $x_2$, replacing the coherent-state projectors~(\ref{eq:Krauscoh}):
\begin{equation}
K_\alpha=\sqrt{\frac{g^2-1}{\pi}}\int\frac{\dtwo\beta}{\pi}\,
e^{-(g-1)|\alpha-\beta|^2}|\beta\rangle\langle\beta|\;.
\end{equation}
This measurement gives up some sensitivity in determining the initial complex amplitude in return for maintaining enough coherence to introduce a bit less noise into the amplified output than does the Arthurs-Kelly measurement~(\ref{eq:Krauscoh}).

\subsubsection{Discussion}

Having surveyed various models for an ideal linear amplifier, we now turn to our main task, formulating a model of phase-preserving linear amplifiers and deriving the complete set of restrictions on the noise that must be added in the amplification process.  Our aim is to draw general conclusions, applicable to any phase-preserving linear amplifier.  The standard input-output relation~(\ref{eq:io}), powerful though it is, is not sufficient for our purpose; the properties of the noise operator $L$ are not sufficiently constrained, beyond the commutator that leads to the second-moment constraint~(\ref{eq:addednoise}), to allow us to draw general conclusions about the full quantum statistics of $L$.  We need a more precise characterization of the operation of a phase-preserving linear amplifier than the input-output relation.  This we give in the next section.

Before moving on, we note that Shi~\textit{et al.}~\cite{ZShi2011a} have considered two models of nonideal amplifiers, a laser amplifier with incomplete inversion and a cascade of alternating ideal amplifiers and ideal attenuators, and found that the nonideal behavior of these amplifiers can be attributed to having an internal noise source that is not in its ground state.  These findings provide additional motivation for our work and are consistent with our general conclusion that any phase-preserving linear amplifier is equivalent to a parametric amplifier with a physical initial state for the ancillary mode, an ideal amplifier arising uniquely in the case where the ancillary mode begins in the vacuum state.

\section{Mathematical characterization of a phase-preserving linear amplifier}
\label{sec:char}

A phase-preserving linear amplifier takes an input signal to an output signal with the same phase, but with amplitude larger by a factor of the amplitude gain $g$.  An essential feature of linearity is that the noise added by the amplifier is independent of the input signal.

In this section, we capture this action mathematically in terms of two superoperators---linear maps on operators---that together characterize the operation of the amplifier.  The first superoperator accounts for the amplification by taking an input coherent state $|\alpha\rangle$ to an output coherent state~$|g\alpha\rangle$:
\begin{equation}
\sA\bigl(|\alpha\rangle\langle\alpha|\bigr)=|g\alpha\rangle\langle g\alpha|\;.
\label{eq:sA}
\end{equation}
The superoperator $\sA$ amplifies without even the amplified input noise and so is clearly not physical by itself.  The second superoperator includes the noise on the output signal by smearing out a phase-space distribution into a broader distribution:
\begin{equation}
\sB=\int\dtwo\beta\,\Pi^{(-1)}(\beta)D(a,\beta)\odot D^\dagger(a,\beta)\;.
\label{sB}
\end{equation}
Here $\odot$ marks the slot where the input to the superoperator goes.  The real-valued function $\Pi^{(-1)}(\beta)$ is assumed to be normalized to unity on the phase plane; we call it the \textit{added-noise function\/} (sometimes the \textit{smearing\/} or \textit{spreading function}).  The added-noise function is independent of the input state, but it can and does depend on the gain~$g$.  We use a superscript $(-1)$ on the added-noise function to indicate that $\Pi^{(-1)}$ has to do with an antinormal ordering, but the connection to antinormal ordering only becomes clear in Sec.~\ref{subsec:model}.  Other orderings for the spreading function will also arise as we proceed.

The overall operation of the amplifier is given by acting first with $\sA$ and then with $\sB$.  This composition of the two is the \textit{amplifier map\/}
\begin{equation}
\sE(\rho)=\sB\bigl(\sA(\rho)\bigr)\equiv\sB\circ\sA(\rho)\;.
\label{eq:sE}
\end{equation}
The natural operator-ordering perspective for the amplifier map becomes apparent when we determine the output state for an input coherent state,
\begin{equation}
\sE\bigl(|\alpha\rangle\langle\alpha|\bigr)
=\int\dtwo\beta\,\Pi^{(-1)}(\beta)D(a,\beta)|g\alpha\rangle\langle g\alpha|D^\dagger(a,\beta)
=\int\dtwo\beta\,\Pi^{(-1)}(\beta-g\alpha)|\beta\rangle\langle\beta|\;.
\end{equation}
For this input, the displaced spreading function, $\Pi^{(-1)}(\beta-g\alpha)$, is the $P$ function of the output state.  Thus the perspective to have in mind is that of the $P$~function in Fig.~\ref{fig1}: $\sA$ amplifies a coherent state without noise, and $\sB$, through the spreading function, accounts for all the noise at the output.

The problem we are interested in can now be expressed as determining the restrictions on the added-noise function $\Pi^{(-1)}$ necessary and sufficient to ensure that the amplifier map $\sE$ can be implemented in a physical system.  Mathematically, this is the requirement that $\sE$ be a completely positive map, i.e., a (trace-preserving) quantum operation.  We have already noted that $\sA$ is not completely positive.  Notice that $\sB$ is a (trace-preserving) quantum operation if $\Pi^{(-1)}(\beta)$ is nonnegative, but we do not need to assume this since it emerges from our main result~\cite{attenuators}.

Strictly speaking, a linear amplifier is phase preserving if and only if $\sE$ commutes with phase-space rotations.  The amplifying map $\sA$ does commute with rotations, but $\sB$ commutes with rotations if and only if $\Pi^{(-1)}(\beta)$ is independent of phase, i.e., depends only on $|\beta|$.  We do not need to assume that $\sB$ has this property, however, to demonstrate our main result.  The only phase-preserving property needed for our main result is built into $\sA$, i.e., that its raw amplification without noise is independent of phase.  Thus we leave $\Pi^{(-1)}(\beta)$ general for the present and make it independent of phase only when we consider examples of nonideal amplifiers in Sec.~\ref{sec:examples} and constraints on the moments of the added noise in~Sec.~\ref{sec:momlim}.

An easily addressed point, which we use as an excuse to introduce a mathematical formulation we need, is that since the coherent-state projectors are not orthogonal, it is not obvious that $\sA$, as defined in Eq.~(\ref{eq:sA}), can be extended by linearity to all operators or, to put it differently, is even a linear map.  To deal with this point, we actually define $\sA$ by its action on the operator basis of displacement operators, which are $\delta$-orthogonal:
\begin{equation}
\tr\bigl(D^\dagger(a,\alpha)D(a,\beta)\bigr)=\pi\delta^2(\alpha-\beta)\;.
\label{eq:Ddeltao}
\end{equation}
Here $\delta^2(\alpha)=\delta(\alpha_R)\delta(\alpha_I)=2\delta(\alpha_1)\delta(\alpha_2)$ is the two-dimensional delta function on the phase plane.  The definition of $\sA$ becomes
\begin{equation}
\sA\bigl(D(a,\beta)\bigr)=\frac{e^{(g^2-1)|\beta|^2/2g^2}}{g^2}D(a,\beta/g)\;.
\label{eq:sAD}
\end{equation}
Linearity and the expression~(\ref{eq:cohFT}) for the coherent-state projectors as a Fourier transform of displacement operators can now be used to \textit{derive\/} Eq.~(\ref{eq:sA}) as the action of the $\sA$ on an input coherent state.

These considerations suggest that it might also be useful to translate the action of $\sB$ to the displacement-operator basis,
\begin{equation}
\sB\bigl(D(a,\beta)\bigr)
=\int\dtwo\alpha\,\Pi^{(-1)}(\alpha)D(a,\alpha)D(a,\beta)D^\dagger(a,\alpha)
=\tilde\Pi^{(-1)*}(\beta)D(a,\beta)\;,
\label{eq:sBD}
\end{equation}
where
\begin{equation}
\tilde\Pi^{(-1)}(\beta)=\int\dtwo\alpha\,\Pi^{(-1)}(\alpha)D(\alpha,\beta)=\tilde\Pi^{(-1)*}(-\beta)
\end{equation}
is the Fourier transform of the added-noise function.  Normalization of the added-noise function is the statement that $\tilde\Pi^{(-1)}(0)=1$.  Notice that these considerations allow us to write the amplifier map as $\sE=\sB\circ\sA=\sA\circ\sB'$, where $\sB'$ is the same as $\sB$ except that its added-noise function is $\tilde\Pi^{(-1)\,\prime}(\beta)=\tilde\Pi^{(-1)}(\beta/g)$ or, equivalently, $\Pi^{(-1)\,\prime}(\alpha)=g^2\Pi^{(-1)}(g\alpha)$.  Since $\sA\circ\sB'$ adds noise first and then does the amplification, this rescaling of the added-noise function is the map version of referring the noise to the input.

Combining Eqs.~(\ref{eq:sAD}) and~(\ref{eq:sBD}) gives the action of $\sE$ in the displacement-operator basis:
\begin{equation}
\sE\bigl(D(a,\beta)\bigr)
=e^{(g^2-1)|\beta|^2/2g^2}\frac{\tilde\Pi^{(-1)*}(\beta/g)}{g^2}D(a,\beta/g)\;.
\label{eq:sED}
\end{equation}
We can write the action~(\ref{eq:sED}) more compactly in terms of antinormally ordered displacement operators, which absorb the Gaussian factors.  A more general approach along these lines is to introduce the $s$-ordering of Cahill and Glauber~\cite{Cahill1969a,Cahill1969b},
\begin{equation}
D^{(s)}(a,\beta)=e^{s|\beta|^2/2}D(a,\beta)\;,
\end{equation}
where $s=0$ gives symmetric ordering of products of $a$ and $a^\dagger$, $s=+1$ gives normal ordering, $D^{(+1)}(a,\beta)=e^{\beta a^\dagger}e^{-\beta^* a}$, and $s=-1$ gives antinormal ordering, $D^{(-1)}(a,\beta)=e^{-\beta^* a}e^{\beta a^\dagger}$.  The $s$-ordered displacement operators satisfy the orthogonality relation
\begin{equation}
\tr\bigl(D^{(s)\dagger}(a,\alpha)D^{(-s)}(a,\beta)\bigr)=\pi\delta^2(\alpha-\beta)\;.
\label{eq:Dsdeltao}
\end{equation}
In terms of $s$-ordering, Eq.~(\ref{eq:sED}) assumes the form
\begin{equation}
\sE\bigl(D^{(s)}(a,\beta)\bigr)=\frac{\tilde\Pi^{(s)*}(\beta/g)}{g^2}D^{(s)}(a,\beta/g)\;,
\label{eq:sEDs}
\end{equation}
where we define an $s$-ordered version of $\tilde\Pi^{(-1)}$,
\begin{equation}
\tilde\Pi^{(s)}(\beta)\equiv
e^{(s+1)(g^2-1)|\beta|^2/2}\tilde\Pi^{(-1)}(\beta)
=e^{s(g^2-1)|\beta|^2/2}\tilde\Pi^{(0)}(\beta)\;,
\end{equation}
which satisfies $\tilde\Pi^{(s)*}(\beta)=\tilde\Pi^{(s)}(-\beta)$ and $\tilde\Pi^{(s)}(0)=1$.  The corresponding (real-valued and normalized) $s$-ordered added-noise function is
\begin{equation}
\Pi^{(s)}(\alpha)=\int\frac{\dtwo\beta}{\pi^2}\,\tilde\Pi^{(s)}(\beta)D(\beta,\alpha)\;,
\end{equation}
although we have no warrant that for $s>-1$, this function is nonnegative or even exists.

Our objective now is, first, to determine how the $s$-ordered characteristic function,
\begin{equation}
\Phi_\rho^{(s)}(\beta)=\bigl\langle D^{(s)}(a,\alpha)\bigr\rangle
=\tr\bigl(\rho D^{(s)}(a,\alpha)\bigr)\;,
\end{equation}
transforms from an input state $\rho$ to the state $\sE(\rho)$ at the output of the linear amplifier and, second, to Fourier transform this result to find the corresponding input-output transformation of the $s$-ordered quasidistribution,
\begin{equation}
W_\rho^{(s)}(\alpha)=\int\frac{\dtwo\beta}{\pi^2}\Phi_\rho^{(s)}(\beta)D(\beta,\alpha)\;,
\end{equation}
where $s=1$ gives the $P$~function, $s=0$ gives the Wigner function, and $s=-1$ gives the $Q$~distribution.  For this purpose, it is useful to translate Eq.~(\ref{eq:sEDs}) to the adjoint $\sE^*$, which is defined by
\begin{equation}
\tr\bigl(A^\dagger\sE(B)\bigr)=\tr\bigl([\sE^*(A)]^\dagger B\bigr)
\end{equation}
and which can be thought of as the Heisenberg-picture version of $\sE$.  Using the $\delta$-orthogonality~(\ref{eq:Dsdeltao}), we have
\begin{equation}
\tr\Bigl(\bigl[\sE^*\bigl(D^{(s)}(a,\alpha)\bigr)\bigr]^\dagger D^{(-s)}(a,\beta)\Bigr)
=\tr\Bigl(D^{(s)\dagger}(a,\alpha)\sE\bigl(D^{(-s)}(a,\beta)\bigr)\Bigr)
=\tilde\Pi^{(-s)*}(\alpha)\pi\delta^2(\beta-g\alpha)\;.
\end{equation}
Comparing the leftmost and rightmost sides of this equality and again using Eq.~(\ref{eq:Dsdeltao}) gives us
\begin{equation}
\sE^*\bigl(D^{(s)}(a,\alpha)\bigr)=\tilde\Pi^{(-s)}(\alpha)D^{(s)}(a,g\alpha)\;.
\label{eq:sEstarD}
\end{equation}

We can now use the action~(\ref{eq:sEstarD}) of $\sE^*$ on displacement operators to determine the input-output transformation of the characteristic function:
\begin{align}
\Phi_{\rm out}^{(s)}(\beta)
&=\tr\bigl(D^{(s)}(a,\beta)\sE(\rho)\bigr)\nonumber\\
&=\tr\Bigl(\bigl[\sE^*\bigl(D^{(s)}(a,-\beta)\bigr)\bigr]^\dagger\rho\Bigr)\nonumber\\
&=\tilde\Pi^{(-s)}(\beta)\tr\bigl(D^{(s)}(a,g\beta)\rho\bigr)\nonumber\\
&=\tilde\Pi^{(-s)}(\beta)\Phi_{\rm in}^{(s)}(g\beta)\;.
\label{eq:Phiio}
\end{align}
The output $s$-ordered quasidistribution is a convolution of the $s$-ordered input quasidistribution, scaled by the gain, with the ($-s$)-ordered spreading function:
\begin{equation}
W_{\rm out}^{(s)}(\alpha)=\int\dtwo\beta\,\Pi^{(-s)}(\alpha-\beta)
\frac{W^{(s)}_{\rm in}(\beta/g)}{g^2}\;.
\label{eq:Wsio}
\end{equation}

The input-output transformation~(\ref{eq:Wsio}) is the generalization of the ball-and-stick depictions in Fig.~\ref{fig1}.  The $P$ function of an input coherent state is a $\delta$ function, so the output $P$ function for this input, as noted above, is given directly by $\Pi^{(-1)}(\alpha)$, displaced to the position of the amplified expectation value of the complex amplitude.  An ideal linear amplifier adds no noise to the $Q$ distribution, so $\Pi^{(+1)}(\alpha)=\delta^2(\alpha)$, which gives the input-output tranformation~(\ref{eq:Qio}) for the $Q$ distribution.  The rescaled Wigner function of an input coherent state $|\gamma\rangle$ is
\begin{equation}
\frac{W_{\rm in}(\beta/g)}{g^2}=\frac{1}{\pi}\frac{e^{-2|\beta-g\gamma|^2/g^2}}{g^2/2}\;;
\end{equation}
for an ideal linear amplifier, this rescaled Wigner function is convolved with
\begin{equation}
\Pi^{(0)}(\alpha-\beta)=\frac{1}{\pi}\frac{e^{-2|\alpha-\beta|^2/(g^2-1)}}{(g^2-1)/2}\;,
\end{equation}
giving an output Wigner function
\begin{equation}
W_{\rm out}(\alpha)=\frac{1}{\pi}\frac{e^{-|\gamma-\beta|^2/(g^2-1/2)}}{g^2-1/2}\;,
\end{equation}
which has the minimum output noise~(\ref{eq:aoutup}) permitted by quantum mechanics.

We now want to go beyond these simple Gaussian considerations and to derive the general constraints that complete positivity of $\sE$ places on the added-noise functions $\Pi^{(s)}$.  A straightforward approach invokes the Kraus representation theorem~\cite{Kraus1983a,Nielsen2000a} to conclude that if $\sE$ is a quantum operation, then there exists an ancilla~$E$, with initial (pure) state $|\phi\rangle\langle\phi|$, and a joint unitary operator $U$ such that
\begin{equation}
\sE(\rho)=\tr_E\bigl(U\rho\otimes|\phi\rangle\langle\phi|U^\dagger\bigr)\;.
\label{eq:sEKraus}
\end{equation}
This is called an \textit{ancilla model\/} for $\sE$ or a \textit{Stinespring extension\/}~\cite{Stinespring1955a}.

It is useful to convert this ancilla model to the adjoint $\sE^*$.  Given any operators $A$ and $B$ on the primary mode, we have
\begin{equation}
\tr_a\bigl(\bigl[\sE^*(A)]^\dagger B)
=\tr_a\bigl(A^\dagger\sE(B)\bigr)
=\tr_{a,E}\bigl(A^\dagger UB\otimes|\phi\rangle\langle\phi|U^\dagger\bigr)
=\tr_a\bigl(\langle\phi\,|U^\dagger A^\dagger U|\phi\rangle B\bigr)
\;,
\end{equation}
which implies, since $B$ is arbitrary, that
\begin{equation}
\sE^*(A)=\langle\phi\,|U^\dagger A U|\phi\rangle\;.
\label{eq:sEstarKraus}
\end{equation}
We can interpret the Kraus representation theorem as saying that $\sE$ is completely positive if and only if there exists a joint unitary and an ancilla state such that $\sE^*$ satisfies Eq.~(\ref{eq:sEstarKraus}) for all operators $A$.

In particular, using Eq.~(\ref{eq:sEstarD}), we have that $\sE$ is completely positive if and only if
\begin{equation}
\tilde\Pi^{(-s)}(\beta)D^{(s)}(a,g\beta)
=\sE^*\bigl(D^{(s)}(a,\beta)\bigr)
=\langle\phi\,|U^\dagger D^{(s)}(a,\beta)U|\phi\rangle\;.
\end{equation}
This expression restricts the way $U$ acts on joint states of the form $|\psi\rangle\otimes|\phi\rangle$, where $|\psi\rangle$ is an arbitrary pure state of the primary mode, and thus seems to promise a way to derive restrictions on the added-noise function, but we have been unable to disentangle those restrictions from the freedom in choosing $U$ and $|\phi\rangle$.  Thus we drop this approach in favor of a different one (leaving open the question of whether the approach based on the Kraus representation theorem can be used to get to our main result).  Instead of relying on the Kraus representation theorem to provide an ancilla model with a joint unitary and a physical ancilla state, we construct an explicit ancilla model using a particular joint unitary, the two-mode squeeze operator, and a ``state'' $\sigma$ of the single ancillary mode.  The ``state'' $\sigma$ determines the added-noise function, and what we prove is that $\sigma$ must be a physical state, i.e., a valid density operator.

\section{Quantum constraints on added-noise functions}
\label{sec:genlim}

\subsection{Two-mode squeezing model for any phase-preserving linear amplifier}
\label{subsec:model}

To develop this second approach, we begin with the $s=0$ version of the input-output transformation~(\ref{eq:Phiio}) for the characteristic function.  We define a unit-trace, Hermitian operator $\sigma$, which we associate with an ancillary mode $b$, by
\begin{equation}
\sigma=\int\frac{\dtwo\beta}{\pi}\,D^\dagger(b,\beta)
\tilde\Pi^{(0)}\bigl(\beta^*/\sqrt{g^2-1}\bigr)\;.
\end{equation}
By the completeness and orthogonality of the displacement operators, this is equivalent to
\begin{equation}
\tilde\Pi^{(0)}(\beta)=\tr_b\bigl(D(b,\sqrt{g^2-1}\,\beta^*)\sigma\bigr)
=\Phi_\sigma^{(0)}\bigl(\sqrt{g^2-1}\,\beta^*\bigr)\;,
\label{eq:FTPizero}
\end{equation}
where $\Phi^{(0)}$ is the symmetrically ordered ``characteristic function'' of the ``state'' $\sigma$.  Quotes are used here because we have no warrant to assume that $\sigma$ is a valid density operator, i.e., has nonnegative eigenvalues.  Indeed, the problem we are interested in, the restrictions on the added-noise function needed to ensure that $\sE$ is completely positive, are translated to the corresponding restrictions on~$\sigma$.  We show in Sec.~\ref{subsec:proof} that $\sigma$ must be a valid state of the ancillary mode~$b$.  Formally, this means that $\sigma$ is a positive operator, having only nonnegative eigenvalues, a property denoted as $\sigma\ge0$.

We can convert the characteristic function~(\ref{eq:FTPizero}) to arbitrary ordering and then Fourier transform to relate the added-noise functions to the corresponding $s$-ordered quasidistributions of~$\sigma$:
\begin{align}
\label{eq:tildePis}
\tilde\Pi^{(s)}(\beta)&=\Phi_\sigma^{(s)}\bigl(\sqrt{g^2-1}\,\beta^*\bigr)\;,\\
\Pi^{(s)}(\alpha)&=\frac{W_\sigma^{(s)}\bigl(-\alpha^*/\sqrt{g^2-1}\bigr)}{g^2-1}\;.
\label{eq:Pis}
\end{align}

These definitions become useful when we introduce the two-mode squeeze operator~(\ref{eq:Sr}) as a joint unitary operator for modes $a$ and $b$; employing the input-output relation~(\ref{eq:iotwo}), we obtain
\begin{align}
\tr_a\Bigl(D(a,\beta)\tr_b\bigl(S\rho\otimes\sigma S^\dagger\bigr)\Bigr)
&=\tr_{a,b}\bigl(S^\dagger D(a,\beta)S\rho\otimes\sigma\bigr)\nonumber\\
&=\tr_a\bigl(D(a,g\beta)\rho\bigr)\tr_b\Bigl(D\bigl(b,\sqrt{g^2-1}\beta^*\bigr)\sigma\Bigr)\nonumber\\
&=\Phi_{\rm in}^{(0)}(g\beta)\tilde\Pi^{(0)}(\beta)\nonumber\\
&=\Phi_{\rm out}^{(0)}(\beta)\nonumber\\
&=\tr_a\bigl(D(a,\beta)\sE(\rho)\bigr)\;.
\end{align}
Since the displacement operators are a complete, $\delta$-orthogonal set of operators,
we get
\begin{equation}
\sE(\rho)=\tr_b\bigl(S\rho\otimes\sigma S^\dagger\bigr)\;,
\end{equation}
just as though we had the ancilla model of Fig.~\ref{fig2}(b), i.e., a single ancillary mode $b$ that interacts with the amplifier mode $a$ via a two-mode squeezing interaction.

\begin{table}
\caption{Linear-amplifier input-output transformations for characteristic functions, $\Phi^{(s)}_{\rm out}$ and $\Phi^{(s)}_{\rm in}$, and quasidistributions, $W^{(s)}_{\rm out}$ and $W^{(s)}_{\rm in}$.  The input-output transformations are specified by added-noise functions, $\tilde\Pi^{(s)}$ or $\Pi^{(s)}$.  $P$, $W$, and $Q$ denote the Glauber-Sudarshan $P$ function, the Wigner $W$ function, and the Husimi $Q$ distribution.  The expressions in square brackets specialize the added-noise-function formula immediately above to an ideal linear amplifier, i.e., to ancillary-mode state $\sigma=|0\rangle\langle0|$.\label{table}}
\begin{ruledtabular}
\begin{tabular}{lcc}
&Characteristic-function transformation
&Quasidistribution transformation\\
$s$ arbitrary
&$\Phi_{\rm out}^{(s)}(\beta)=\tilde\Pi^{(-s)}(\beta)\Phi_{\rm in}^{(s)}(g\beta)$
&$\displaystyle{W_{\rm out}^{(s)}(\alpha)=\int\dtwo\beta\,\Pi^{(-s)}(\alpha-\beta)
\frac{W^{(s)}_{\rm in}(\beta/g)}{g^2}}$\\[8pt]
{}
&$\tilde\Pi^{(-s)}(\beta)=\Phi_\sigma^{(-s)}\bigl(\sqrt{g^2-1}\beta^*\bigr)$
&$\displaystyle{\Pi^{(-s)}(\alpha)=\frac{W_\sigma^{(-s)}\bigl(-\alpha^*/\sqrt{g^2-1}\bigr)}{g^2-1}}$\\[8pt]
{}
&$\Bigl[e^{-(1+s)(g^2-1)|\beta|^2/2}\Bigr]$
&$\displaystyle{\biggl[\frac{1}{\pi}\frac{e^{-2|\alpha|^2/(1+s)(g^2-1)}}{(1+s)(g^2-1)/2}\biggr]}$\\[14pt]
$s=+1$
&$\Phi_{\rm out}^{(+1)}(\beta)=\tilde\Pi^{(-1)}(\beta)\Phi_{\rm in}^{(+1)}(g\beta)$
&$\displaystyle{P_{\rm out}(\alpha)=\int\dtwo\beta\,\Pi^{(-1)}(\alpha-\beta)
\frac{P_{\rm in}(\beta/g)}{g^2}}$\\[6pt]
&$\tilde\Pi^{(-1)}(\beta)=\Phi_\sigma^{(-1)}\bigl(\sqrt{g^2-1}\beta^*\bigr)$
&$\displaystyle{\Pi^{(-1)}(\alpha)=\frac{Q_\sigma\bigl(-\alpha^*/\sqrt{g^2-1}\bigr)}{g^2-1}}$\\[8pt]
{}
&$\Bigl[e^{-(g^2-1)|\beta|^2}\Bigr]$
&$\displaystyle{\biggl[\frac{1}{\pi}\frac{e^{-|\alpha|^2/(g^2-1)}}{g^2-1}\biggr]}$\\[14pt]
$s=0$
&$\Phi_{\rm out}^{(0)}(\beta)=\tilde\Pi^{(0)}(\beta)\Phi_{\rm in}^{(0)}(g\beta)$
&$\displaystyle{W_{\rm out}(\alpha)=\int\dtwo\beta\,\Pi^{(0)}(\alpha-\beta)
\frac{W_{\rm in}(\beta/g)}{g^2}}$\\[6pt]
&$\tilde\Pi^{(0)}(\beta)=\Phi_\sigma^{(0)}\bigl(\sqrt{g^2-1}\beta^*\bigr)$
&$\displaystyle{\Pi^{(0)}(\alpha)=\frac{W_\sigma\bigl(-\alpha^*/\sqrt{g^2-1}\bigr)}{g^2-1}}$\\[8pt]
{}
&$\Bigl[e^{-(g^2-1)|\beta|^2/2}\Bigr]$
&$\displaystyle{\biggl[\frac{1}{\pi}\frac{e^{-2|\alpha|^2/(g^2-1)}}{(g^2-1)/2}\biggr]}$\\[14pt]
$s=-1$
&$\Phi_{\rm out}^{(-1)}(\beta)=\tilde\Pi^{(+1)}(\beta)\Phi_{\rm in}^{(-1)}(g\beta)$
&$\displaystyle{Q_{\rm out}(\alpha)=\int\dtwo\beta\,\Pi^{(+1)}(\alpha-\beta)
\frac{Q_{\rm in}(\beta/g)}{g^2}}$\\[6pt]
&$\tilde\Pi^{(+1)}(\beta)=\Phi_\sigma^{(+1)}\bigl(\sqrt{g^2-1}\beta^*\bigr)$
&$\displaystyle{\Pi^{(+1)}(\alpha)=\frac{P_\sigma\bigl(-\alpha^*/\sqrt{g^2-1}\bigr)}{g^2-1}}$\\[8pt]
{}
&$\bigl[1\bigr]$
&$\displaystyle{\bigl[\delta^2(\alpha)\bigr]}$\\[8pt]
\end{tabular}
\end{ruledtabular}
\end{table}

It is trivial that if $\sigma\ge0$, then $\sE$ is completely positive, so what we have to prove is that if $\sE$ is completely positive, then $\sigma\ge0$.  Jiang, Piani, and Caves~\cite{Jiang2012a} have established the necessary and sufficient properties of a unitary operator $U$ such that the complete positivity of a map
\begin{equation}
\sE(\rho)=\tr_E(U\rho\otimes\sigma U^\dagger)
\label{eq:sEmodel}
\end{equation}
implies that the ancilla ``state'' $\sigma$ is a valid density operator.  They show that a sufficient, but not necessary condition on $U$ is that it be full rank, i.e., that the ancilla operators in its operator Schmidt decomposition span the space of operators on the ancilla.  The two-mode squeeze operator is full rank, so we could rely on the results of~\cite{Jiang2012a} to assert our main result.  Since the proof in~\cite{Jiang2012a} assumes finite dimensions, however, we first prove, in Sec.~\ref{subsec:proof}, that $S(r)$ is full rank for $r\ne0$ and then use this result to show that $\sigma\ge0$.

Our main result thus is that \textit{any\/} phase-preserving linear amplifier is equivalent to a parametric amplifier with a physical state $\sigma$ for the ancillary mode.  An ideal linear amplifier is the case where $\sigma$ is the vacuum state.  Stated in terms of the added-noise functions, our result shows that they are rescaled quasidistributions of the state $\sigma$.  In particular, for our formulation of a linear amplifier in terms of the maps $\sA$ and $\sB$, which uses the $P$-function perspective of Fig.~\ref{fig1}, the added-noise function $\Pi^{(-1)}$ is a rescaled $Q$ distribution of the ancillary mode; the added-noise function $\Pi^{(-1)}$ is thus required to be everywhere nonnegative, although this is by no means sufficient to guarantee that the added-noise function is physical.  If, instead, we use the $Q$-distribution perspective of Fig.~\ref{fig1}, the added-noise function $\Pi^{(+1)}$ is a rescaled $P$ function of the ancillary mode.  Finally, if we use the symmetrically ordered moments that are usually used to discuss amplifier noise, the added-noise function $\Pi^{(0)}$ is a rescaled Wigner function of the ancillary mode.  These relations are summarized in Table~\ref{table}.

We also note that for an ideal linear amplifier, the input-output transformation can be written as $\Phi_{\rm out}^{(s)}(\beta)=\Phi_{\rm in}^{(s')}(g\beta)$ or, equivalently, as $W_{\rm out}^{(s)}(\alpha)=W_{\rm in}^{(s')}(\alpha/g)/g^2$, where $s'=s-(1+s)(1-1/g^2)$.  If, for example, $s=1$, then $s'=1-2(1-1/g^2)$, implying that in the limit of high gain, the output $P$ function looks like a rescaled input $Q$ distribution.  For arbitrary input states, as $g\rightarrow\infty$, the amplifier noise wipes out any singularities or negativity associated with the input $P$ function.  Nonetheless, Nha, Milburn, and Carmichael~\cite{Nha2010a} have shown that there are input states for which nonclassical features persist in the output for arbitrarily large gain.

\subsection{Proof of main result}
\label{subsec:proof}

To prove our main result, we first establish that the squeeze operator is full rank; i.e., we show that given an operator $O_b$ on the ancillary mode, $\tr_b(SO_b)=0$ implies $O_b=0$.  We begin by writing the two-mode squeeze operator in a form similar to that in Eq.~(\ref{eq:Sfactored}):
\begin{equation}
S(r)
=g^{-(a^\dagger a+b^\dagger b+1)/2}
e^{-a^\dagger b^\dagger\sqrt{g^2-1}}e^{ab\sqrt{g^2-1}}
g^{-(a^\dagger a+b^\dagger b+1)/2}\;.
\end{equation}
This gives us, for any coherent state $|\alpha\rangle$ on mode~$a$,
\begin{align}
\langle\alpha|S(r)|\alpha\rangle
&=e^{-|\alpha|^2(1-1/g)}
g^{-(b^\dagger b+1)/2}
\langle\alpha/\sqrt g|e^{-a^\dagger b^\dagger\sqrt{g^2-1}}
e^{ab\sqrt{g^2-1}}|\alpha/\sqrt g\rangle
g^{-(b^\dagger b+1)/2}\nonumber\\
&=e^{|\alpha|^2(g-1)^2/2g}
g^{-(b^\dagger b+1)/2}
D\biggl(b,-\sqrt{\frac{g^2-1}{g}}\alpha^*\biggr)
g^{-(b^\dagger b+1)/2}\;.
\end{align}
Our premise, that $\tr_b(SO_b)=0$, implies that for all $\alpha$,
\begin{equation}
0=\tr_b\bigl(\langle\alpha|S|\alpha\rangle O_b\bigr)=
e^{|\alpha|^2(g-1)^2/2g}
\tr_b\!\left(
D\biggl(b,-\sqrt{\frac{g^2-1}{g}}\alpha^*\biggr)
g^{-(b^\dagger b+1)/2}O_b g^{-(b^\dagger b+1)/2}
\right)\;.
\end{equation}
Since the displacement operators are a complete, $\delta$-orthogonal set of operators, we get that $g^{-(b^\dagger b+1)/2}O_b g^{-(b^\dagger b+1)/2}=0$ and, hence, by the invertibility of $g^{-(b^\dagger b+1)/2}$, that $O_b=0$.  This establishes that $S$ is full rank.

Suppose now that we have a full-rank joint unitary $U$ and a quantum operation defined as in Eq.~(\ref{eq:sEmodel}).  Diagonalize the ancilla initial ``state'' $\sigma$, and decompose it into positive-eigenvalue and negative-eigenvalue parts,
\begin{equation}
\sigma=\sigma_+-\sigma_-\;,
\end{equation}
where
\begin{align}
\sigma_+&=\sum_{j+}\lambda_{j+}|e_{j+}\rangle\langle e_{j+}|\;,\\
\sigma_-&=\sum_{j-}\lambda_{j-}|e_{j-}\rangle\langle e_{j-}|\;.
\end{align}
The eigenvectors $|e_{j\pm}\rangle$ make up an orthonormal basis.  We assume that the $\lambda_{j+}$s are strictly positive and allow the $\lambda_{j-}$s to be positive or zero (zero so as to fill out the orthonormal basis with zero-eigenvalue eigenvectors).  If we take the ancilla trace in a basis $|f_k\rangle$, the quantum operation takes the form
\begin{equation}
\sE(\rho)
=\sum_{k,j+}M_{k,j+}\rho M_{k,j+}^\dagger-
\sum_{k,j-}M_{k,j-}\rho M_{k,j-}^\dagger\;,
\end{equation}
where
\begin{equation}
M_{k,j\pm}=\sqrt{\lambda_{j\pm}}\langle f_k|U|e_{j\pm}\rangle
=\sqrt{\lambda_{j\pm}}\tr_b\bigl(U|e_{j\pm}\rangle\langle f_k|\bigr)
\label{eq:Mkj}
\end{equation}
are ``operation elements'' that decompose $\sE$ into positive and negative parts.

If any nonzero negative-part operation element, say $M_{K,J-}$, does not lie in the operator subspace spanned by the positive-part operation elements, $\{M_{k,j+}\}$, then by projecting $M_{K,J-}$ orthogonal to this operator subspace, we obtain a nonzero operator $N_{J,K-}$ such that
\begin{align}
\sum_{k,j+}\bigl|\tr\bigl(N_{J,K-}^\dagger M_{j,k+}\bigr)\bigr|^2
-\sum_{k,j-}\bigl|\tr\bigl(N_{J,K-}^\dagger M_{j,k-}\bigr)\bigr|^2
&=-\sum_{k,j-}\bigl|\tr\bigl(N_{J,K-}^\dagger M_{j,k-}\bigr)\bigr|^2\cr
&\le-\bigl|\tr\bigl(N_{J,K-}^\dagger M_{J,K-}\bigr)\bigr|^2\cr
&<0\;,
\end{align}
which implies that $\sE$ is not completely positive (for a completely positive $\sE$, which thus has a Kraus decomposition, there can be no operator like $N_{J,K-}$).  Thus, no matter what $U$ is, complete positivity of $\sE$ requires that all the negative-part operation elements lie in the span of the positive-part operation elements.

This means that for any $K$ and $J-$,
\begin{equation}
M_{K,J-}=\sum_{k,j+}c^{K,J-}_{k,j+}M_{k,j+}\;,
\end{equation}
for some coefficients $c^{K,J-}_{k,j+}$.  Using the definition~(\ref{eq:Mkj}), we can rewrite this expression as
\begin{equation}
0=
\tr_b\Biggl(U
\biggl(\sqrt{\lambda_{J-}}|e_{J-}\rangle\langle f_K|
-\sum_{k,j+}c^{K,J-}_{k,j+}\sqrt{\lambda_{j+}}|e_{j+}\rangle\langle f_k|\biggr)
\Biggr)\;.
\end{equation}
That $U$ is full rank implies that
\begin{equation}
\sqrt{\lambda_{J-}}|e_{J-}\rangle\langle f_K|
-\sum_{k,j+}c^{K,J-}_{k,j+}\sqrt{\lambda_{j+}}|e_{j+}\rangle\langle f_k|
=0\;.
\label{eq:fr}
\end{equation}
Applying this expression to $|f_l\rangle$, we get
\begin{equation}
0=\sqrt{\lambda_{J-}}|e_{J-}\rangle\delta_{Kl}-\sum_{j+}c^{K,j-}_{l,j+}\sqrt{\lambda_{j+}}|e_{j+}\rangle\;.
\end{equation}
Since the vectors $|e_{j\pm}\rangle$ are linearly independent, all the coefficients in this expression must be zero.  In particular, for $l=K$, we get that $\lambda_{J-}=0$.  Since this holds for all $J-$, we can conclude that $\sigma_-=0$.  Thus, we reach the desired conclusion that for a full-rank $U$, $S(r)$ being an example, $\sigma$ must be a valid density operator.

That $\sigma$ must be a positive operator has been proven for the amplifier transformation and for more general phase-space transformations~\cite{Demoen1977a}, using a technique that does not reveal the connection to a particular unitary operator, the two-mode squeeze operator in the case of a linear amplifier.

\section{Examples of nonideal linear amplifiers}
\label{sec:examples}

We reiterate that the key assumption necessary for our proof is the phase-preserving amplification carried out by the map $\sA$.  The proof does not require that the added-noise function be phase insensitive, i.e., that $\sigma$ be invariant under phase-space rotations, nor even that $\sigma$ have zero mean complex amplitude.  Since phase-preserving amplifiers do have phase-insensitive noise, however, we assume that $\sigma$ is rotationally invariant for the examples considered in this section and for the discussion in Sec.~\ref{sec:momlim} of quantum limits on moments of the added noise.  Rotational invariance implies that $\sigma$ is diagonal in the number basis
\begin{equation}\label{eq:sigman}
\sigma=\sum_{n=0}^\infty\lambda_n|n\rangle\langle n|\;.
\end{equation}
In this section we consider both physical and unphysical $\sigma$ of this form, thus allowing us to compare physical and unphysical linear amplifiers.

For rotationally invariant $\sigma$, the added-noise number~(\ref{eq:addednoiseno}) is given by
\begin{equation}
A_1=\langle|\,b\,|^2\rangle=\langle b^\dagger b\rangle_\sigma+\frac{1}{2}\;.
\end{equation}
The second-moment quantum limit, $A_1\ge1/2$, becomes the constraint that $\sigma$ have a nonnegative mean number of quanta, $\langle b^\dagger b\rangle_\sigma=\langle n\rangle_\sigma\ge0$.

The $s$-ordered characteristic function for a number state~\cite{Cavesnotes} is
\begin{equation}
\Phi_n^{(s)}(\beta)=e^{s|\beta|^2/2}\langle n|D(b,\beta)|n\rangle=
e^{(s-1)|\beta|^2/2}L_n\bigl(|\beta|^2\bigr)\;,
\end{equation}
where $L_n$ denotes the Laguerre polynomial of degree~$n$.  Fourier transforming gives the corresponding quasidistribution for a number state~\cite{Cavesnotes},
\begin{equation}
W_n^{(s)}(\alpha)=\int\frac{\dtwo\beta}{\pi^2}\Phi_n^{(s)}(\beta)D(\beta,\alpha)
=\frac{2}{\pi(1-s)}\left(\frac{s+1}{s-1}\right)^n
e^{-2|\beta|^2/(1-s)}L_n\biggl(\frac{4}{1-s^2}|\beta|^2\biggr)\;.
\end{equation}
We can plug these results into Eqs.~(\ref{eq:tildePis}) and (\ref{eq:Pis}) to find series representations of the added-noise functions, $\tilde\Pi^{(s)}$ and $\Pi^{(s)}$, for the general rotationally invariant state~(\ref{eq:sigman}).  The series representation is particularly useful when $\sigma$ has only a few nonzero eigenvalues.

\begin{figure}[htbp]
{\ifnum\figstyle=0 \includegraphics[width=\textwidth]{fig5.eps} \else\includegraphics[width=\textwidth]{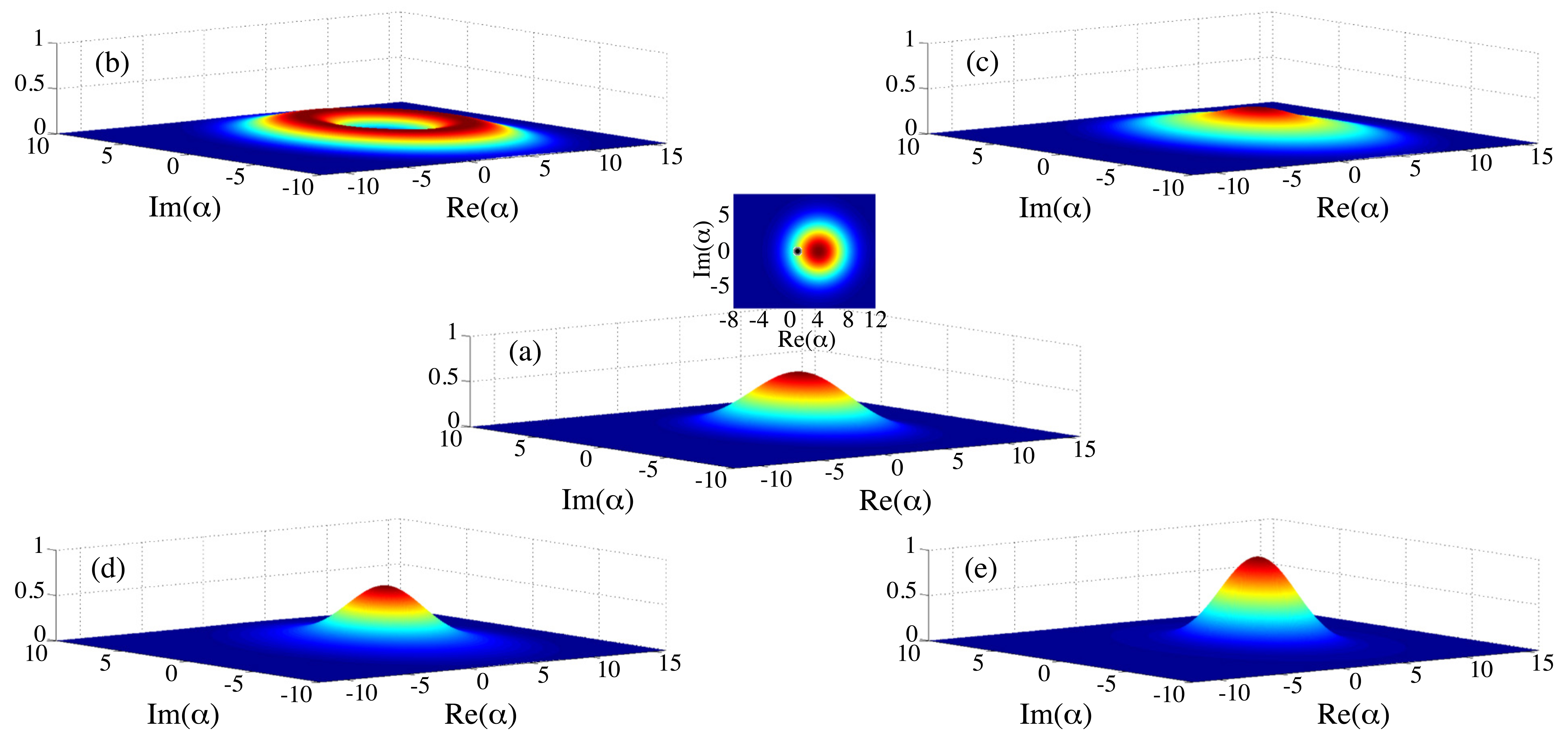} \fi}
\caption{Output $P$ functions for five amplifiers.  For all five, the input is a coherent state $|\beta\rangle$, with $\beta=1$, and the amplitude gain is $g=4$; these correspond to the conditions in Figs.~\protect{\ref{fig0}} and~\protect{\ref{fig1}}.  The output $P$ function is given by the added-noise function as in Eq.~(\protect{\ref{eq:Pout}}).  In (a) the ancillary-mode state $\sigma$ is the vacuum state, which gives an ideal linear amplifier.  The inset shows the $P$ function of the input coherent state as a tiny black dot and the output $P$ function, $P_{\rm out}(\alpha)=e^{-|\alpha|^2/(g^2-1)}/\pi(g^2-1)$, as colored contours; this inset is the full contour plot of the $P$-function perspective that is given as a stick figure in Fig.~\protect{\ref{fig1}}.  The main plot in~(a) displays the same output $P$ function in a three-dimensional plot.  In (b)--(e), the ancillary-mode ``state'' is $\sigma=(\frac{1}{2}-\lambda)|0\rangle\langle0|+\lambda|1\rangle\langle1|+\frac{1}{2}|2\rangle\langle2|$ [Eq.~(\protect{\ref{eq:sigma3}})]: (b)~$\lambda=0.5$, (c)~$\lambda=0$, (d)~$\lambda=-0.5$, and~(e)~$\lambda=-1.0$; the output $P$ functions, given by Eqs.~(\protect{\ref{eq:Pout}}) and~(\protect{\ref{eq:Piminus3}}), are displayed as three-dimensional plots.  Plots~(b) and~(c) are physical amplifiers, whereas (d) and (e) are unphysical, this despite the similarity, at least by eye, of (d) to the Gaussian of the ideal linear amplifier in (a).  All four values of $\lambda$ are such that the second-moment quantum limit is satisfied, which requires $\lambda\ge-1$, and the output $P$ function is everywhere nonnegative, which requires $-1.37=-(1+\sqrt3)/2\le\lambda\le1/2$.}
\label{fig5}
\end{figure}

In this subsection, we use the $P$-function perspective introduced in Secs.~\ref{subsec:ideal} and~\ref{sec:char}, for which we need the $Q$ function of a number state,
\begin{equation}
Q_n(\alpha)=\frac{1}{\pi}|\langle\alpha|n\rangle|^2
=\frac{e^{-|\alpha|^2}}{\pi}\frac{|\alpha|^{2n}}{n!}\;.
\end{equation}
The added-noise function for the rotationally invariant state~(\ref{eq:sigman}) is
\begin{equation}\label{eq:Piminus}
\Pi^{(-1)}(\alpha)=\sum_{n=0}^\infty\lambda_n\frac{Q_n\bigl(-\alpha^*/\sqrt{g^2-1}\bigr)}{g^2-1}
=\frac{e^{-|\alpha|^2/(g^2-1)}}{\pi(g^2-1)}
\sum_{n=0}^\infty\lambda_n\frac{|\alpha|^{2n}}{n!(g^2-1)^n}\;.
\end{equation}

To illustrate the possibilities for the added noise, we specialize now to a one-parameter family of states, which have support only on the first three number states,
\begin{align}\label{eq:sigma3}
\sigma
=\lambda_0|0\rangle\langle0|+\lambda_1|1\rangle\langle1|+\lambda_2|2\rangle\langle2|
=(\textstyle{\frac{1}{2}}-\lambda)|0\rangle\langle0|
+\lambda|1\rangle\langle1|+\textstyle{\frac{1}{2}}|2\rangle\langle2|\;.
\end{align}
For this $\sigma$ to be physical, the three eigenvalues must be nonnegative, which restricts the parameter $\lambda$ to the range $0\le\lambda\le\frac{1}{2}$; here we will also be considering values outside this range, which give us unphysical $\sigma$ and, hence, unphysical amplifiers. The mean number of quanta, $\langle b^\dagger b\rangle_\sigma=\lambda_1+2\lambda_2=\lambda+1$, gives a second-moment constraint, $\lambda\ge-1$; that this second-moment quantum limit allows unphysical values of $\lambda$ indicates that quantum constraints on higher moments are important.

\begin{figure}[htbp]
{\ifnum\figstyle=0 \includegraphics[width=\textwidth]{fig6.eps} \else\includegraphics[width=\textwidth]{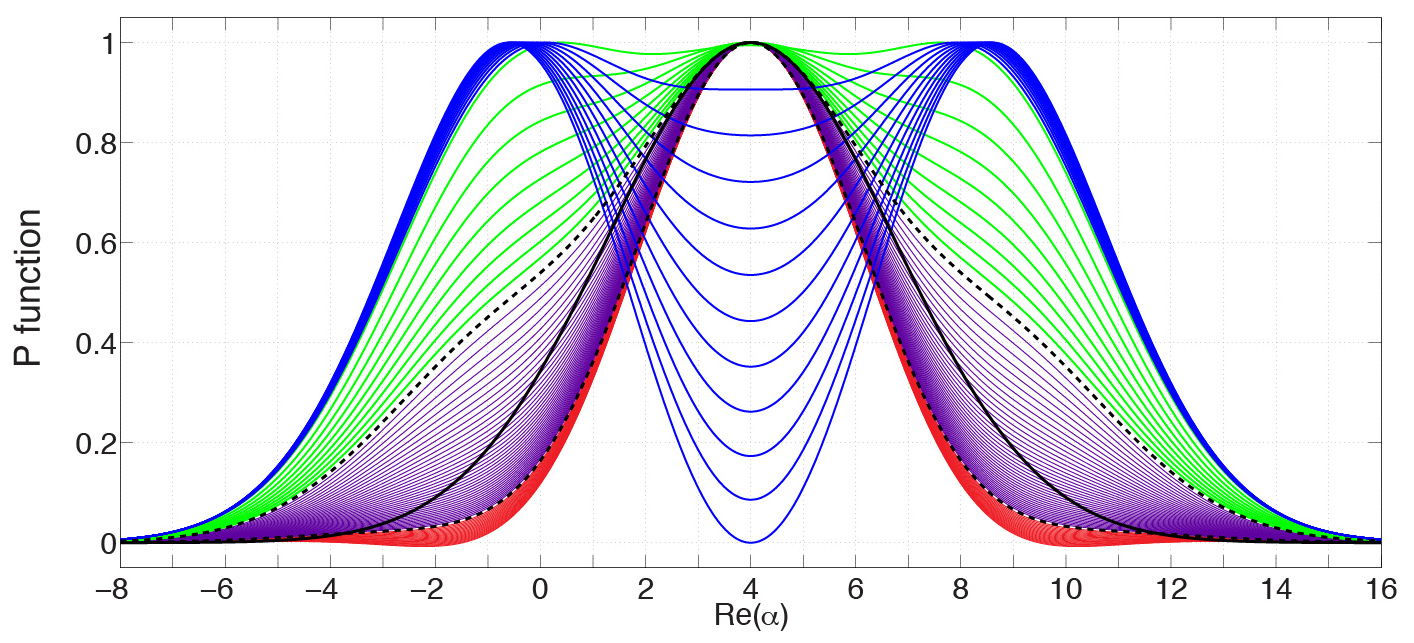} \fi}
\caption{Output $P$ function~(\protect{\ref{eq:Pout}}), which is a displacement of the added-noise function~(\protect{\ref{eq:Piminus3}}), for amplifiers with ancillary-mode ``state'' $\sigma$ of Eq.~(\protect{\ref{eq:sigma3}}).  The parameter $\lambda$ varies from $0.5$ to $-1.5$ in steps of size $0.025$.  As in Fig.~\ref{fig3}, the input state is a coherent state $|\beta\rangle$ with $\beta=1$, and the amplitude gain is $g=4$.  The $P$ function is plotted along the real $\alpha$ axis, which is sufficient because of the rotational symmetry of the noise.  The plotted $P$ functions are not normalized to unity; instead they are scaled to have maximum value equal to~1. \  The amplifiers in the range $0<\lambda\le0.5$, all of which are physical, are plotted in green and blue; $\lambda=0$, the boundary between physical and unphysical, is plotted as a black dashed line.  When $\lambda=0.5$, there is no vacuum contribution to $\sigma$, so the $P$ function has a deep hole in the middle.  As $\lambda$ decreases from $0.5$, the vacuum contribution increases, and the hole disappears, to be replaced by a distribution with broad shoulders, still faintly evident at $\lambda=0$.  The transition from distributions that have a minimum in the middle (blue) to those with a maximum in the middle (green) occurs at $\lambda=0.22$.  The values in the range $-1<\lambda<0$, in which the second-moment constraint is satisfied ($A_1\ge1/2$), are plotted in purple; $\lambda=-1$ is plotted as a black dashed line.  The values $\lambda<-1$, plotted in red, violate the second-moment constraint, and for $\lambda<-(1+\sqrt3)/2=-1.37$, the output $P$ function takes on negative values.  The Gaussian output $P$ function of an ideal linear amplifier is plotted as a solid line; it reveals the nonGaussian character of all the added-noise functions associated with the ancillary-mode ``state''  $\sigma$ of Eq.~(\protect{\ref{eq:sigma3}}).}
\label{fig6}
\end{figure}

In our examples, we assume that the input to the amplifier is a coherent state $|\beta\rangle$.  In this situation, the output $P$ function is given by the added-noise function~(\ref{eq:Piminus}), displaced to the amplified mean complex amplitude, i.e.,
\begin{equation}\label{eq:Pout}
P_{\rm out}(\alpha)=\Pi^{(-1)}(\alpha-g\beta)\;.
\end{equation}
We can ask for the range of values of $\lambda$ for which the added-noise function,
\begin{equation}\label{eq:Piminus3}
\Pi^{(-1)}(\alpha)
=\frac{e^{-|\alpha|^2/(g^2-1)}}{\pi (g^2-1)}
\biggl(\lambda_0
+\lambda_1\frac{|\alpha|^2}{g^2-1}
+\frac{1}{2}\lambda_2\frac{|\alpha|^4}{(g^2-1)^2}\biggr)\;,
\end{equation}
is everywhere nonnegative.  We need $\lambda_0\ge0$ to keep the added-noise function nonnegative at $\alpha=0$, and we need $\lambda_2\ge0$ to keep it nonnegative when $|\alpha|^2$ is large (if $\lambda_2=0$, we also need $\lambda_1\ge0$).  When $\lambda_2>0$, the polynomial $\lambda_0+\lambda_1 x+\frac{1}{2}\lambda_2x^2$, which multiplies the Gaussian in the added-noise function, with $x=|\alpha|^2/(g^2-1)$, has a minimum at $x=-\lambda_1/\lambda_2$, where it takes on the value $\lambda_0-\lambda_1^2/2\lambda_2$.  We don't care about this minimum when $\lambda_1>0$, since the minimum then occurs at a negative value of $x$, but when $\lambda_1\le0$, ensuring that the added-noise function is nonnegative requires that $\lambda_0-\lambda_1^2/2\lambda_2\ge0$, i.e., $2\lambda_0\lambda_2\ge\lambda_1^2$.  The requirements for nonnegativity of the added-noise function are thus (i)~$\lambda_0\ge0$, (ii)~$\lambda_2\ge0$, and (iii)~$2\lambda_0\lambda_2\ge\lambda_1^2$ if $\lambda_1\le0$.  Translated to the single parameter $\lambda$, the requirements for the nonnegativity of the added-noise function reduce to $-(1+\sqrt3)/2\le\lambda\le1/2$.

Figure~\ref{fig5} displays the output $P$ function~(\ref{eq:Pout}), which is a displacement of the added-noise function~(\ref{eq:Piminus3}), for four values of $\lambda$, two physical and two unphysical, using the ancillary-mode state~(\ref{eq:sigma3}).  For comparison, the figure also shows the output $P$ function for an ideal linear amplifier ($\lambda_0=1$, $\sigma=|0\rangle\langle0|$) with the same input state and the same gain.  The lessons to be drawn from Fig.~\ref{fig5} are, first, that eyeballing the added-noise function is not a reliable way to assess its physicality and, second, that the second-moment quantum limit is not sufficient to discriminate physical from unphysical amplifiers, since all four values of $\lambda$ satisfy the second-moment constraint.  Figure~\ref{fig6} provides a more detailed look at the output $P$ function for values of $\lambda$ between $0.5$ and $-1.5$.  The Gaussian output of an ideal linear amplifier is shown for comparison, making it easy to see the nonGaussian character of the output noise.  Even so, it is not easy to judge whether these added-noise functions correspond to physical amplifiers just by looking at the plots, except for those values, $\lambda<-(1+\sqrt3)/2$, for which the output $P$ function goes negative.

\section{Quantum limits on added-noise moments}
\label{sec:momlim}

In this section we turn our principal result into quantum constraints on the moments of the added noise.  We assume that all the noise, both the noise carried by the input signal and the added noise, is phase insensitive; for the added noise, this implies that $\sigma$ has the form~(\ref{eq:sigman}).  The only nonzero moments of $\Delta a=a-\langle a\rangle$, which we call \textit{noise moments}, are those for which the number of creation operators matches the number of annihilation operators.

We characterize the noise in terms of symmetrically ordered noise moments.  For this purpose we introduce the notation $|\,a\,|^{2k}$ for the symmetric product of $k$ annihilation operators and $k$ creation operators; formally, we can write
\begin{equation}
|\,a\,|^{2k}=\left.
\frac{\partial^{2k}D(a,\alpha)}{\partial\alpha^k\partial(-\alpha^*)^k}
\,\right|_{\alpha=\alpha^*=0}=
\frac{(k!)^2}{(2k)!}
\begin{pmatrix}
\mbox{sum of the $(2k)!/(k!)^2$ ways of ordering}\\
\mbox{a product of $k$ annihilation operators}\\
\mbox{and $k$ creation operators}
\end{pmatrix}\;.
\label{eq:symprod}
\end{equation}
In terms of this notation, the nonvanishing input and output noise moments are written as $\langle|\Delta a|^{2k}\rangle$.  Details of our moment manipulations are relegated to an Appendix.

The input-output relation for the nonvanishing noise moments is
\begin{equation}
\bigl\langle|\Delta a_{\rm out}|^{2k}\bigr\rangle
=\sum_{m=0}^k
\left(\frac{k!}{m!(k-m)!}\right)^{\!2}g^{2(k-m)}(g^2-1)^m
\bigl\langle|\Delta a_{\rm in}|^{2(k-m)}\bigr\rangle A_m\;,
\label{eq:momentioAk}
\end{equation}
where
\begin{equation}
A_k\equiv\bigl\langle|\,b\,|^{2k}\bigr\rangle_\sigma\;,
\label{eq:Ak}
\end{equation}
the $(2k)$th symmetric moment of $b$ and $b^\dagger$, we call the \textit{$k$th added-noise number}.  Equation~(\ref{eq:momentioAk}) expresses how the the input noise, given by the moments $\bigl\langle|\Delta a_{\rm in}|^{2(k-m)}\bigr\rangle$, combines with the added noise, given by the added noise numbers $A_m$, to produce an output noise moment.  The last ($m=k$) term in the sum~(\ref{eq:momentioAk}) comes only from the added-noise number $A_k$; it characterizes the noise added at the $(2k)$th moment.  If the signal noise is known \textit{a priori}, then the added-noise numbers can be obtained from measurements of successive output noise moments; such a procedure has been implemented, using dual input and output ports, in~\cite{Menzel2010a}.

Using Eq.~(\ref{eq:normtosym}), the added-noise numbers $A_k$ can be written in terms of normally ordered moments, or \textit{factorial moments}, $\langle(b^\dagger)^m b^m\rangle_\sigma=\langle(b^\dagger b)(b^\dagger b-1)\cdots(b^\dagger b-m+1)\rangle_\sigma$,
\begin{equation}
A_k=\frac{k!}{2^k}\sum_{m=0}^k\frac{k!}{m!(k-m)!}\frac{2^m}{m!}\langle(b^\dagger)^m b^m\rangle_\sigma
\;.
\end{equation}
Since all the terms in this sum are nonnegative, $A_k$ is bounded below by the state-independent $m=0$ term, which is equal to $k!/2^k$.  This bound is achieved if and only if all the higher terms in the sum vanish, and that occurs if and only if $\sigma$ is the vacuum state.  Thus we have the following quantum limits on the added-noise numbers:
\begin{equation}
A_k\ge\bigl\langle0\bigl||\,b\,|^{2k}\bigr|0\bigr\rangle=\frac{k!}{2^k}\;.
\label{eq:Akqlim}
\end{equation}
For comparison, a thermal state,
\begin{equation}
\sigma=\frac{1}{1+\bar n}
\left(\frac{\bar n}{1+\bar n}\right)^{b^\dagger b}\;,
\end{equation}
with mean number of quanta $\bar n$, has factorial moments $\langle(b^\dagger)^k b^k\rangle=k!\,\bar n^k$ and added-noise numbers $A_k=k!\bigl(\bar n+\frac{1}{2}\bigr)^k$.

The quantum limits~(\ref{eq:Akqlim}) are not very useful, except for the familiar second-moment constraint $A_1\ge1/2$.  One way to see this is to return to the family of ``states'' of Eq.~(\ref{eq:sigma3}).  For any rotationally invariant $\sigma$, as in Eq.~(\ref{eq:sigman}), for which $\lambda_n=0$ for $n>N$, the factorial moments $\langle(b^\dagger)^kb^k\rangle_\sigma$ vanish for $k>N$.  For the states of Eq.~(\ref{eq:sigma3}), all the factorial moments vanish except $\langle b^\dagger b\rangle_\sigma=\lambda_1+2\lambda_2=\lambda+1$ and $\langle(b^\dagger)^2b^2\rangle_\sigma=2\lambda_2=1$, and these give added-noise numbers $A_k=(k!/2^k)\bigl(1+2k[\lambda_1+(k+1)\lambda_2]\bigr)$.  The quantum limit~(\ref{eq:Akqlim}) is satisfied if $\lambda_1\ge-(k+1)\lambda_2$ [$\lambda\ge-(k+1)/2$].  Thus the ``states'' with $\lambda\ge-1$ satisfy the quantum limits~(\ref{eq:Akqlim}) for all $k$; in particular, the two unphysical ``states'' depicted in Fig.~\ref{fig5} satisfy all these quantum limits.

To do better, we need conditions on the noise that are necessary and sufficient to guarantee that $\sigma$ is a valid density operator.  Since we specialize to phase-insensitive added noise, for which $\sigma$ is diagonal in the number basis, we are dealing with a classical probability distribution, defined by the eigenvalues $\lambda_n$.  Thus the appropriate conditions can be obtained from the solution of the classical moment problem~\cite{Akhiezer1965a,Christiansen2004a,Reed1972a}: what sequences of moments are consistent with a (necessarily nonnegative) probability distribution?  The answer to this question depends on the domain on which the probability distribution is defined: the Hamburger moment problem deals with a distribution defined on the entire real line, i.e., $-\infty<x<\infty$; the Hausdorff moment problem deals with a distribution defined on a finite interval, which can be taken to be $0\le x\le 1$; neatly sandwiched between these two problems, the Stieltjes moment problem deals with the domain $0\le x<\infty$ and thus provides the meat to feed into the maw of our analysis.

The moments in our situation are moments of the number operator:
\begin{equation}
M_l\equiv\langle(b^\dagger b)^l\rangle_\sigma=\langle n^l\rangle_\sigma\;.
\label{eq:numbermom}
\end{equation}
The Stieltjes problem is stated in terms of a sequence $M_l$, $l=1,2,\ldots\,$, from which one constructs two sequences of matrices,
\begin{align}
Q_k^{(0)}&=
\begin{pmatrix}
1&M_1&M_2&\cdots&M_k\\
M_1&M_2&M_3&\cdots&M_{k+1} \\
M_2&M_3&M_4&\cdots&M_{k+2}\\
\vdots&\vdots&\vdots&\ddots\\
M_k&M_{k+1}&M_{k+2}&\cdots&M_{2k}
\end{pmatrix}\;,\\[8pt]
Q_k^{(1)}&=
\begin{pmatrix}
M_1&M_2&M_3&\cdots&M_{k+1}\\
M_2&M_3&M_4&\cdots&M_{k+2} \\
M_3&M_4&M_5&\cdots&M_{k+3}\\
\vdots&\vdots&\vdots&\ddots\\
M_{k+1}&M_{k+2}&M_{k+3}&\cdots&M_{2k+1}
\end{pmatrix}\;,
\end{align}
$k=0,1,2,\ldots\,$.  The sequence $M_l$ consists of moments of a nonnegative probability distribution if and only if (i)~$\det Q_k^{(0)}>0$ and $\det Q_k^{(1)}>0$ for $k=0,1,2,\ldots\,$ or (ii)~$\det Q_k^{(0)}>0$ and $\det Q_k^{(1)}>0$ for $k=0,1,\ldots,K$ and $\det Q_k^{(0)}=0$ and $\det Q_k^{(1)}=0$ for $k>K$.  Case~(i) leads to a distribution with infinite support, and case~(ii) to a distribution with finite support.  These then are the amplifier quantum limits expressed in terms of number moments.  In the following, we give these quantum limits as in case~(i), i.e., with strict inequalities, but discuss the consequences of case~(ii) for equalities in the quantum limits.  We take no account of the fact that we are concerned with distributions concentrated on the nonnegative integers, whereas the Stieltjes problem deals with distributions on the continuous domain $0\le x<\infty$.

The first four nontrivial quantum limits imposed by the solution to the Stieltjes problem are the following:
\begin{subequations}
\label{eq:Mlim}
\begin{align}
0&< M_1\;,\label{eq:M1}\\
0&< M_2-M_1^2\;,\label{eq:M2}\\
0&< M_1M_3-M_2^2\;,\\
0&< M_4(M_2-M_1^2)-M_3(M_3-M_1M_2)+M_2(M_1M_3-M_2^2)\;.
\end{align}
\end{subequations}
The consequence of case~(ii) is that there can be equalities in this list, but once one encounters an equality, all subsequent constraints in the list must also be equalities.

The first constraint~(\ref{eq:M1}) is simply that the mean number of quanta is positive.  Allowing for equality in this constraint gives the usual second-moment quantum limit.  Equality implies that $\sigma$ is the vacuum state (thus an ideal linear amplifier); since all higher moments also vanish, all the constraints become equalities.  The second constraint~(\ref{eq:M2}) requires that the variance of the number of quanta be positive.  The consequence of equality in this constraint is that the variance is zero, which implies that $\sigma$ is a number eigenstate and thus that all the constraints except the first are equalities.  Notice that the first three constraints imply $0<M_1(M_3-M_1M_2)$, which since $M_1>0$, implies $0<M_3-M_1M_2$.

We can apply the number-moment quantum limits~(\ref{eq:Mlim}) to the one-parameter family of ``states''~(\ref{eq:sigma3}), for which the number moments are $M_l=\lambda_1+2^l\lambda_2=\lambda+2^{l-1}$.  The first four quantum limits reduce to $\lambda>-1$; $\lambda^2+\lambda-1<0$, i.e., $-(1+\sqrt5)/2<\lambda<(-1+\sqrt5)/2$; $\lambda>0$; and $\lambda(\frac{1}{2}-\lambda)>0$.  The first two quantum limits do not rule out all unphysical states, the third rules out unphysical states with negative values of $\lambda$, and the fourth rules out all unphysical states.  One cannot achieve equality in the first two constraints, since doing so is unphysical, but the third and fourth constraints can achieve equality.

To be able to use the number-moment quantum limits, we need the relations between number moments and the added-noise numbers.  Using Eq.~(\ref{eq:prodtosym}), the added-noise numbers can be written in terms of number moments as
\begin{equation}
A_k=\sum_{l=0}^k M_l
\frac{k!}{2^k}\sum_{m=l}^k\frac{k!}{m!(k-m)!}\frac{2^m}{m!}S_m^{(l)}\;,
\label{eq:prodtosymAk}
\end{equation}
where $S_k^{(l)}$ denotes a (signed) \textit{Stirling number of the first kind\/}~\cite{Abramowitz1964a,wiki}. The first four cases of Eq.~(\ref{eq:prodtosymAk}) are the following:
\begin{subequations}
\begin{align}
A_1&=M_1+\textstyle{\frac{1}{2}}\;,\\
A_2&=M_2+M_1+\textstyle{\frac{1}{2}}\;,\\
A_3
&=M_3+\textstyle{\frac{3}{2}}M_2+2M_1+\textstyle{\frac{3}{4}}\;,\\
A_4
&=M_4+2M_3+5M_2+4M_1+\textstyle{\frac{3}{2}}\;.
\end{align}
\end{subequations}
The constant term in these expressions is the quantum limit~(\ref{eq:Akqlim}).  More useful is to write the number moments in terms of added-noise numbers.  The number moments can be written in terms of factorial moments using Eq.~(\ref{eq:normtoprod}) and in terms of the added-noise numbers using Eq.~(\ref{eq:symtoprod}):
\begin{equation}
M_l=
\sum_{k=0}^l\mathcal{S}_l^{(k)}\langle(b^\dagger)^k b^k\rangle_\sigma
=\sum_{m=0}^l A_m
\sum_{k=m}^l\!\left(-\frac{1}{2}\right)^{k-m}\frac{k!}{m!}\frac{k!}{m!(k-m)!}\mathcal{S}_l^{(k)}\;,
\label{eq:symtoprodMl}
\end{equation}
Here $\mathcal{S}_l^{(k)}$, defined in Eq.~(\ref{eq:Stirling2}), denotes a \textit{Stirling number of the second kind\/}~\cite{Abramowitz1964a,wiki}.  The first four cases of Eq.~(\ref{eq:symtoprodMl}) are the following:
\begin{subequations}
\begin{align}
M_1&=A_1-\textstyle{\frac{1}{2}}\;,\\
M_2&=A_2-A_1\;,\\
M_3&=A_3-\textstyle{\frac{3}{2}}A_2-\textstyle{\frac{1}{2}}A_1+\textstyle{\frac{1}{4}}\;,\\
M_4&=A_4-2A_3-2A_2+2A_1\;.
\end{align}
\end{subequations}
Plugging these expressions into the number-moment quantum limits~(\ref{eq:Mlim}) gives the first four quantum limits in terms of the added-noise numbers:
\begin{subequations}
\label{eq:Akqlim2}
\begin{align}
A_1&>\frac{1}{2}\;,\\
A_2&>A_1^2+\frac{1}{4}\;,\\
A_3&>\frac{1}{2}\!\left(3A_2+A_1-\frac{1}{2}\right)+\frac{(A_2-A_1)^2}{A_1-\frac{1}{2}}\;,\\
A_4&>2(A_3+A_2-A_1)\nonumber\\
&\phantom{=\;}+
\frac{(A_2-A_1)^3+\frac{1}{16}(4A_3-6A_2-2A_1+1)[8A_1(A_1-A_2)+4A_3-2A_2-6A_1+1]}
{A_2-A_1^2-\frac{1}{4}}\;.
\end{align}
\end{subequations}
The complexity of the last of these expressions suggests that the best way to deal with quantum limits on higher moments of the added noise is to translate measured added-noise numbers into effective number moments using Eq.~(\ref{eq:symtoprodMl}) and to use the quantum limits expressed in terms of number moments, as in Eqs.~(\ref{eq:Mlim}).

\section{Conclusion}
\label{sec:con}

Amplification, by translating from the real world of quantum physics to the mundane world of everyday experience, is a principal means by which we gain access to the quantum world.  Phase-preserving amplification transforms signals too weak to be perceived into much larger signals that we can lay our grubby, classical hands on.  As the signal transitions to the classical world, however, quantum mechanics extracts its due: any phase-preserving linear amplifier must add noise, which is equivalent to half a quantum at the input in the limit of high gain.

In this paper we consider the full set of quantum constraints on the operation of a single-mode phase-preserving linear amplifier, going well beyond the usual emphasis on second moments and Gaussian noise.  Our main result is that any phase-preserving linear amplifier is equivalent to a parametric amplifier with a single ancillary mode that begins in a physical state $\sigma$.  The noise properties of the amplifier, even should it bear no resemblance to a parametric amplifier, are encoded in the effective state~$\sigma$.  In particular, the noise the amplifier adds to a signal, as encoded in symmetrically ordered moments, is described completely by the Wigner function of $\sigma$.  Using this general characterization of linear amplification, we consider how the phase-space quasidistributions of the noise input to the amplifier and the noise added by amplification combine to produce the noise at the output of the amplifier, and we derive quantum limits on all moments of the added noise.

Despite the length of this paper, there is work still to be done.  Perhaps the most important extension of our work will be to amplification of continuous-time signals.  Such signals are best dealt with in the Fourier domain, where we can think of a phase-preserving linear amplifier as one that amplifies a continuum of frequency modes with a frequency-dependent gain.  We expect our main result to generalize in the obvious way: at each frequency, the amplifier will be equivalent to a parametric amplifier, with an ancillary mode that provides the frequency-dependent gain at that frequency; the joint state of all the ancillary modes will have to be physical, but the ancillary modes will not have to be independent, even in the case of time-stationary noise.  The second moments of the added noise will be expressed in terms of a spectral density of added noise, which will obey the usual quantum limit~\cite{Caves1982a,Clerk2004a}.  For Gaussian noise, the added-noise spectral density will be the entire story, but for a nonGaussian amplifier, there will not only be the possibility of nonGaussian ancillary-mode states, as in a single-mode amplifier, but also the possibility of entanglement among the ancillary modes at different frequencies.

A second extension involves how best to characterize the performance of a phase-preserving linear amplifier.  We derive in Sec.~\ref{sec:momlim} the quantum limits on the measured moments of the added noise.  These limits are both cumbersomely complicated and not really the point.  The best way to characterize the performance of a linear amplifier would be to translate the measured noise into an effective ancillary-mode state $\sigma$, i.e., into estimates of the eigenvalues $\lambda_n$ of $\sigma$.  A nearly quantum-limited amplifier, for example, will have $\lambda_0$ close to 1.  Thus what one would like to do is to perform \textit{indirect tomography\/}~\cite{Jiang2012a}, in which one uses measurements on a system, in this case the amplified modes, to reconstruct the state of (perhaps imaginary) ancillas, in this case the ancillary modes of our parametric-amplifier model.  In the amplifier context, this sort of tomography is a species of optical homodyne or heterodyne tomography~\cite{Lvovsky2009a,Kiukas2010a}, since one uses the statistics of linear measurements at the output of the amplifier, like homodyne or heterodyne measurements, to reconstruct a quantum state, in this case the state of the imaginary ancillary modes.  This sort of tomography is a tricky business, fraught with instabilities.  We defer consideration of it to future work, not just because it's tricky, although that is a problem, but also because the job really should be done in the context of continuous-time, multi-frequency linear amplification.

\begin{acknowledgments}
The authors thank S.~D. Bartlett and G.~J. Milburn for helpful and enlightening conversations.  We thank N.~C. Menicucci for participating actively in an extended, critical ``Fuchsian analysis'' of an initial draft of part of this paper.  We thank C.~A. Fuchs for originating and lending his name to---the dubbing was done by S.~T. Flammia---this excruciatingly thorough, joint reading of a manuscript by all its co-authors.  Fuchsian analysis is a procedure we recommend, tedious though it is, to authors of all scientific papers.  This work was supported in part by National Science Foundation Grant Nos.~PHY-0903953, PHY-1212445, and PHY-1005540 and by Office of Naval Research Grant No.~N00014-11-1-0082.
\end{acknowledgments}

\appendix*

\section{Symmetrically and normally ordered products and number powers}

In this Appendix, we give the relations among ordered products and powers of the number operator that are used in Sec.~\ref{sec:momlim}.  For details on the hypergeometric function ${}_2F_1$, see~\cite{Abramowitz1964a}, and for details on the Stirling numbers, see~\cite{Abramowitz1964a,wiki}.

The input-output relation for a phase-preserving linear amplifier, expressed in terms of
symmetrically ordered characteristic functions ($s=0$), combines Eqs.~(\ref{eq:Phiio}) and~(\ref{eq:tildePis}) into
\begin{equation}\label{eq:Phiio2}
\Phi_{\rm out}^{(0)}(\beta)
=\Phi_{\rm in}^{(0)}(g\beta)\Phi_\sigma^{(0)}\bigl(\sqrt{g^2-1}\,\beta^*\bigr)\;.
\end{equation}
If we assume that $\sigma$ has no mean field, as is true for the rotationally invariant $\sigma$ of Eq.~(\ref{eq:sigman}), the expectation value of the complex amplitude of the primary mode transforms as in Eq.~(\ref{eq:expio}).  Factoring out the input and output expectation values from the characteristic functions gives new characteristic functions,
\begin{align}
\Psi_{\rm in}^{(0)}(\beta)
&=e^{-\beta\langle a_{\rm in}\rangle^*+\beta^*\langle a_{\rm in}\rangle}
\Phi_{\rm in}^{(0)}(\beta)=\langle D(\Delta a_{\rm in},\beta)\rangle\;,\\
\Psi_{\rm out}^{(0)}(\beta)
&=e^{-\beta\langle a_{\rm out}\rangle^*+\beta^*\langle a_{\rm out}\rangle}
\Phi_{\rm out}^{(0)}(\beta)=\langle D(\Delta a_{\rm out},\beta)\rangle\;,
\end{align}
which generate symmetrically ordered noise moments, i.e., moments of $\Delta a=a-\langle a\rangle$.  In terms of these new characteristic functions, the input-output relation~(\ref{eq:Phiio2}) becomes
\begin{equation}\label{eq:Psiio}
\Psi_{\rm out}^{(0)}(\beta)
=\Psi_{\rm in}^{(0)}(g\beta)\Phi_\sigma^{(0)}\bigl(\sqrt{g^2-1}\,\beta^*\bigr)\;.
\end{equation}
If we further assume that all the noise is phase insensitive, i.e., that the characteristic functions in Eq.~(\ref{eq:Psiio}) depend only on $|\beta|$, then the only nonzero moments are those with an equal number of creation and annihilation operators.  The input-output relation for these noise moments is
\begin{align}
\bigl\langle|\Delta a_{\rm out}|^{2k}\bigr\rangle
&=\left.
\frac{\partial^{2k}\Psi_{\rm out}^{(0)}(\beta)}{\partial\beta^k\partial(-\beta^*)^k}
\,\right|_{\beta=\beta^*=0}\nonumber\\
&=\sum_{m=0}^k
\left(\frac{k!}{m!(k-m)!}\right)^{\!2}
\left.\frac{\partial^{2(k-m)}\Psi_{\rm in}^{(0)}(g\beta)}
{\partial\beta^{k-m}\,\partial(-\beta^*)^{k-m}}\right|_{\beta=\beta^*0}
\left.\frac{\partial^{2m}\Phi_\sigma^{(0)}\bigl(\sqrt{g^2-1}\beta\bigr)}
{\partial\beta^m\,\partial(-\beta^*)^m}\right|_{\beta=\beta^*=0}\nonumber\\
&=\sum_{m=0}^k
\left(\frac{k!}{m!(k-m)!}\right)^{\!2}g^{2(k-m)}(g^2-1)^m
\bigl\langle|\Delta a_{\rm in}|^{2(k-m)}\bigr\rangle
\bigl\langle|\,b\,|^{2m}\bigr\rangle_\sigma\;.
\label{eq:momentio}
\end{align}
In the second line, all the other possible derivatives vanish as a consequence of phase insensitivity, i.e., because the characteristic functions depend only on the absolute value
of their arguments.  The last term in the sum ($m=k$) comes only from the added noise and characterizes the noise added at the $(2k)$th moment, so we define it, in Eq.~(\ref{eq:Ak}), to be the $k$th added-noise number.

To derive quantum limits on the added-noise numbers, we need to relate them to moments of the number operator $b^\dagger b$.  We do this in two steps.  The symmetrically ordered product~(\ref{eq:symprod}) is related to normally ordered products by
\begin{align}
|\,b\,|^{2k}
&=
\left.
\frac{\partial^{2k}[e^{-\beta\beta^*/2}D^{(+1)}(b,\beta)]}{\partial\beta^k\,\partial(-\beta^*)^k}
\,\right|_{\beta=\beta^*=0}\nonumber\\
&=\sum_{m=0}^k
\left(\frac{k!}{m!(k-m)!}\right)^{\!2}
\left.
\frac{\partial^{2(k-m)}e^{-\beta\beta^*/2}}
{\partial\beta^{(k-m)}\,\partial(-\beta^*)^{(k-m)}}
\right|_{\beta=\beta^*=0}
\left.
\frac{\partial^{2m}D^{(+1)}(b,\beta)}
{\partial\beta^m\,\partial(-\beta^*)^m}
\right|_{\beta=\beta^*=0}
\nonumber\\
&=\frac{k!}{2^k}\sum_{m=0}^k\frac{k!}{m!(k-m)!}\frac{2^m}{m!}
(b^\dagger)^m b^m
\;.
\label{eq:normtosym}
\end{align}
Notice that $\langle n|(b^\dagger)^kb^k|n\rangle=n(n-1)\cdots(n-k+1)=(-1)^k(-n)_k=$, where $(x)_k=x(x+1)\cdots(x+k-1)=(x+k-1)!/(x-1)!$, $k=0,1,2,\ldots\,$, denotes the \textit{Pochhammer symbol\/} (or \textit{rising factorial}).  The \textit{falling factorial\/} can be written in terms of the Pochhammer symbol as $(-1)^k(-x)_k=x(x-1)\cdots(x-k+1)=x!/(x-k)!\,$.  In terms of this notation, the normally ordered product,
\begin{equation}
(b^\dagger)^kb^k=b^\dagger b\,(b^\dagger b-1)\cdots(b^\dagger b-k+1)=(-1)^k(-b^\dagger b)_k\;,
\end{equation}
is the falling factorial.

The second step is to write the normally ordered products in terms of powers of the number operator.  One way to do this is to iterate the recursion relation,
\begin{equation}
(b^\dagger b)_k
=(b^\dagger b-k+1)(b^\dagger b)_{k-1}\;,
\end{equation}
to generate the required relations, the first four of which are
\begin{subequations}
\begin{align}
b^\dagger b&=b^\dagger b\;,\\
(b^\dagger)^2b^2&=(b^\dagger b)^2-b^\dagger b\;,\\
(b^\dagger)^3b^3&=(b^\dagger b)^3-3(b^\dagger b)^2+2b^\dagger b\;,\\
(b^\dagger)^4b^4&=(b^\dagger b)^4-6(b^\dagger b)^3+11(b^\dagger b)^2-6b^\dagger b\;.
\end{align}
\end{subequations}
Generally, we can use the expansion of the falling factorial as a polynomial in $x$,
\begin{equation}
(-1)^k(-x)_k=\sum_{l=0}^k S_k^{(l)}x^l\;,
\label{eq:PochStirling}
\end{equation}
where the coefficients $S_k^{(l)}$, $l\le k=0,1,\ldots\,$, are the (signed) \textit{Stirling numbers of the first kind}.  The Stirling numbers satisfy $S_k^{(0)}=\delta_{k0}$, which makes $(x)_0=1$.   Equation~(\ref{eq:PochStirling}) converts normally ordered products to powers of the number operator:
\begin{equation}
(b^\dagger)^k b^k=(-1)^k(-b^\dagger b)_k=\sum_{l=0}^k S_k^{(l)}(b^\dagger b)^l\;.
\label{eq:prodtonorm}
\end{equation}

Using the Pochhammer symbol, we can rewrite Eq.~(\ref{eq:normtosym}) as
\begin{equation}
|\,b\,|^{2k}
=\frac{k!}{2^k}
\sum_{m=0}^k
\frac{(-b^\dagger b)_m(-k)_m}{(1)_m}\frac{2^m}{m!}
=\frac{k!}{2^k}\,{}_2F_1(-b^\dagger b,-k;1;2)
=\frac{k!}{2^k}(-1)^{b^\dagger b}\,{}_2F_1(-b^\dagger b,k+1;1;2)\;,
\label{eq:normtosym2}
\end{equation}
where ${}_2F_1$ denotes the hypergeometric function.  Plugging Eq.~(\ref{eq:prodtonorm}) into Eq.~(\ref{eq:normtosym}) gives us a closed-form expression for symmetric products in terms of powers of the number operator:
\begin{equation}
|\,b\,|^{2k}=\sum_{l=0}^k(b^\dagger b)^l
\frac{k!}{2^k}\sum_{m=l}^k\frac{k!}{m!(k-m)!}\frac{2^m}{m!}S_m^{(l)}\;.
\label{eq:prodtosym}
\end{equation}
The first four cases are the following:
\begin{subequations}
\begin{align}
|\,b\,|^2&=b^\dagger b+\textstyle{\frac{1}{2}}\;,\\
|\,b\,|^4&=(b^\dagger b)^2+b^\dagger b+\textstyle{\frac{1}{2}}\;,\\
|\,b\,|^6
&=(b^\dagger b)^3+\textstyle{\frac{3}{2}}(b^\dagger b)^2+2b^\dagger b+\textstyle{\frac{3}{4}}\;,\\
|\,b\,|^8
&=(b^\dagger b)^4+2(b^\dagger b)^3+5(b^\dagger b)^2+4b^\dagger b+\textstyle{\frac{3}{2}}\;. 
\end{align}
\end{subequations}
The last term in each expression is equal to $k!/2^k$.

We can invert Eq.~(\ref{eq:prodtosym}) by following the same steps in the opposite direction.
The normally ordered products are related to symmetric products by
\begin{align}
(b^\dagger)^k b^k
&=
\left.
\frac{\partial^{2k}[e^{\beta\beta^*/2}D(b,\beta)]}{\partial\beta^k\,\partial(-\beta^*)^k}
\,\right|_{\beta=\beta^*=0}\nonumber\\
&=\sum_{m=0}^k
\left(\frac{k!}{m!(k-m)!}\right)^{\!2}
\left.
\frac{\partial^{2(k-m)}e^{\beta\beta^*/2}}
{\partial\beta^{(k-m)}\,\partial(-\beta^*)^{(k-m)}}
\right|_{\beta=\beta^*=0}
\left.
\frac{\partial^{2m}D(b,\beta)}
{\partial\beta^m\,\partial(-\beta^*)^m}
\right|_{\beta=\beta^*=0}
\nonumber\\
&=\frac{(-1)^k k!}{2^k}\sum_{m=0}^k\frac{k!}{m!(k-m)!}\frac{(-2)^m}{m!}
|\,b\,|^{2m}
\;.
\label{eq:symtonorm}
\end{align}
The \textit{Stirling numbers of the second kind},
\begin{equation}
\mathcal{S}_l^{(k)}=\frac{1}{k!}
\sum_{m=0}^k(-1)^{k-m}\frac{k!}{m!(k-m)!}m^l\;,\quad k\le l=0,1,\ldots\;,
\label{eq:Stirling2}
\end{equation}
are the matrix inverse of the Stirling numbers of the first kind, i.e.,
\begin{equation}
\delta_{ll'}=\sum_{k=l}^{l'}S_k^{(l)}\mathcal{S}_{l'}^{(k)}
=\sum_{k=l}^{l'}S_{l'}^{(k)}\mathcal{S}_k^{(l)}\;.
\end{equation}
This can be used to invert Eq.~(\ref{eq:PochStirling}),
\begin{equation}
x^l=\sum_{k=0}^l\mathcal{S}_l^{(k)}(-1)^k(-x)_k\;,
\label{eq:PochStirling2}
\end{equation}
and, hence, to invert the corresponding operator relation~(\ref{eq:prodtonorm}):
\begin{equation}
(b^\dagger b)^l=\sum_{k=0}^l\mathcal{S}_l^{(k)}(b^\dagger)^k b^k\;.
\label{eq:normtoprod}
\end{equation}
Now, plugging Eq.~(\ref{eq:symtonorm}) into Eq.~(\ref{eq:normtoprod}) gives us
\begin{equation}
(b^\dagger b)^l=\sum_{m=0}^l|\,b\,|^{2m}
\sum_{k=m}^l\!\left(-\frac{1}{2}\right)^{k-m}\frac{k!}{m!}\frac{k!}{m!(k-m)!}\mathcal{S}_l^{(k)}\;,
\label{eq:symtoprod}
\end{equation}
of which the first four cases are the following:
\begin{subequations}
\begin{align}
b^\dagger b&=|\,b|^2-\textstyle{\frac{1}{2}}\;,\\
(b^\dagger b)^2&=|\,b\,|^4-|\,b\,|^2\;,\\
(b^\dagger b)^3
&=|\,b\,|^6-\textstyle{\frac{3}{2}}|\,b\,|^4-\textstyle{\frac{1}{2}}|\,b\,|^2+\textstyle{\frac{1}{4}}\;,\\
(b^\dagger b)^4
&=|\,b\,|^8-2|\,b\,|^6-2|\,b\,|^4+2|\,b\,|^2\;.
\end{align}
\end{subequations}

\end{document}